\documentclass[12pt,letterpaper]{article}

\usepackage{natbib,hyperref}

\usepackage[ left=1in, top=1in, right=1in, bottom=1in]{geometry}
\usepackage{graphicx,bm,colonequals,amsmath,amssymb,url,xcolor,bbm}
\usepackage{array,tabularx,multirow}
\usepackage[font={footnotesize}]{caption,subcaption}
\usepackage[utf8]{inputenc}
\usepackage{enumitem}
\usepackage{soul} %
\usepackage{placeins} %
\usepackage[normalem]{ulem}
\usepackage{minted}

\graphicspath{{plots/}}

\usepackage{xr}
\makeatletter
\newcommand*{\addFileDependency}[1]{%
  \typeout{(#1)}
  \@addtofilelist{#1}
  \IfFileExists{#1}{}{\typeout{No file #1.}}
}
\makeatother
\newcommand*{\myexternaldocument}[1]{%
    \externaldocument{#1}%
    \addFileDependency{#1.tex}%
    \addFileDependency{#1.aux}%
}
\myexternaldocument{supplement}

\usepackage{mathtools}

\everydisplay{\textstyle}

\bibpunct[, ]{(}{)}{;}{a}{,}{,}
   
\usepackage{amsthm}
\newtheoremstyle{propstyle} %
    {2mm}                    %
    {1mm}                    %
    {\itshape}                   %
    {}                           %
    {\scshape}                   %
    {.}                          %
    {.5em}                       %
    {}  %
\theoremstyle{propstyle}

\theoremstyle{propstyle}

\theoremstyle{propstyle}

\theoremstyle{propstyle}

\theoremstyle{propstyle}

\usepackage[ruled,vlined]{algorithm2e}
\SetKwInput{KwInput}{Input}
\usepackage{algorithmic}
\usepackage{array,framed}
\usepackage{float}

\makeatletter
\renewcommand{\paragraph}{%
  \@startsection{paragraph}{4}%
  {\z@}{2ex \@plus 1ex \@minus .2ex}{-1em}%
  {\normalfont\normalsize\bfseries}%
}
\makeatother

\DeclareMathAlphabet\mathbfcal{OMS}{cmsy}{b}{n}

\newcommand{\bb}{\mathbf{b}}
\newcommand{\bs}{\mathbf{s}}

\newcommand{\bx}{\mathbf{x}}
\newcommand{\by}{\mathbf{y}}

\newcommand{\bz}{\mathbf{z}}

\newcommand{\bA}{\mathbf{A}}

\newcommand{\bG}{\mathbf{G}}

\newcommand{\bI}{\mathbf{I}}

\newcommand{\bK}{\mathbf{K}}

\newcommand{\bQ}{\mathbf{Q}}

\newcommand{\bfzero}{\mathbf{0}}

\newcommand{\bftheta}{\bm{\theta}}

\newcommand{\GP}{\mathcal{GP}}

\newcommand{\normal}{\mathcal{N}}

\newcommand{\locs}{\mathcal{S}}

\title{Generative multi-scale modeling via spatial autoregressive transport maps}

\author{Alejandro Calle-Saldarriaga\thanks{Department of Statistics, University of Wisconsin--Madison}\footnotemark[1] \and Paul F.V.\ Wiemann\thanks{Department of Statistics, The Ohio State University} \and Matthias Katzfuss\footnotemark[1] \thanks{Corresponding author: \texttt{katzfuss@gmail.com}}}

\date{}

\begin{document}

\maketitle

\begin{abstract}
Spatial fields in the Earth and environmental sciences are often available at multiple scales or resolutions. While coarse-scale data (e.g., from global circulation models) are often abundant, they lack the local detail provided by fine-scale data (e.g., from regional climate models), which are typically computationally expensive to generate. Statistical downscaling and multi-scale data fusion address this challenge by predicting high-resolution fields from low-resolution or related inputs. We propose a highly scalable Bayesian approach that can learn the joint non-Gaussian distribution and nonlinear dependence structure of nonstationary spatial fields across multiple scales from a small number of training samples. Our method employs scale-aware autoregressive Gaussian processes with suitably chosen regularization-inducing priors to model the conditional distribution of fine-scale fields given coarse-scale data. Exploiting conjugacy, the integrated likelihood is available in closed form, enabling efficient parameter optimization via stochastic gradient descent. Once trained, the method provides a closed-form characterization of the posterior distribution of fine-scale fields given coarse-scale inputs. In numerical comparisons, we demonstrate that our approach substantially outperforms existing methods and effectively characterizes and simulates fine-scale climate behavior based on output from coarse global circulation models.
\end{abstract}


\section{Introduction}\label{sec:intro}

\paragraph{Downscaling and multi-scale fusion}

\begin{figure}[tpb]
    \centering
    \includegraphics[width=\linewidth]{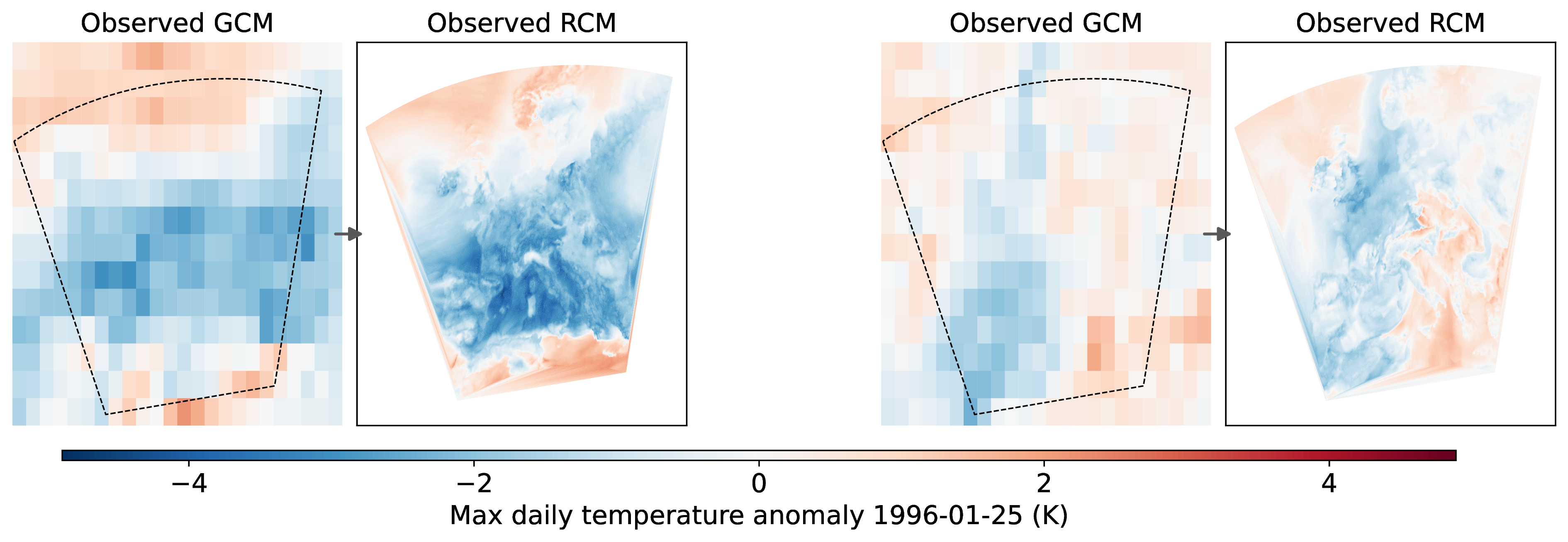}
    \caption{Our goal is to learn the non-Gaussian joint and conditional distributions of spatial fields at multiple scales from a small number of training samples. Here we show an example of two pairs of low-resolution GCM temperature fields at $N_1 = 336$ coarse pixels over Europe driving high-resolution RCM samples on a fine grid of size $N_2 = 280 \times 280 = 78{,}400$; see Section~\ref{sec:climate} for more details. We want to learn the conditional $N_2$-dimensional distribution given a $N_1$-dimensional field from $n \leq 40$ training sample pairs.}
    \label{fig:anomalies_samples}
\end{figure}

Spatial fields in the Earth and environmental sciences are often available from multiple sources with varying resolutions, reliabilities, and physical variables. A primary example involves climate data: Global circulation models (GCMs) provide projections of future climates but are often too spatially coarse to resolve fine-scale processes with crucial local impacts \citep[e.g.][]{Mearns2009}. Dynamical downscaling of these coarse models using regional climate models (RCMs) provides high-resolution details but is computationally expensive \citep{giorgi1989, coppola2021, giorgi2023}. Figure \ref{fig:anomalies_samples} shows samples of GCM-RCM pairs. 

This challenge is not limited to climate models. Other examples include fusing low-resolution satellite soil moisture data with sparse high-resolution ground station data, or coupling meteorology-chemistry models that require higher resolutions in urban areas \citep[e.g.,][]{singh2023deep,zhang2012urban}. In these settings, ``coarse-scale'' data are often abundant, while ``fine-scale'' or high-quality data are scarce or expensive to generate. Integrating such heterogeneous data sources, which may describe the same or related underlying quantities at different resolutions, is a fundamental challenge in multi-scale data fusion \citep[e.g.,][]{chou1994multiscale, ramasamy2013fusion,chen2009regional} and statistical downscaling \citep[e.g.,][]{Berrocal2010,gonzalez2023multi}.

\paragraph{The need for stochastic surrogates}

To alleviate the computational burden of dynamical downscaling or high-resolution data collection, statistical downscaling methods often construct a stochastic surrogate for the costlier process \citep{mearns2017}. While many standard methods focus on point predictions, there is a crucial need for probabilistic approaches that capture the full distribution of the fine-scale field \citep{maraun2018statistical,maraun2010downscaling, pan2025mesmer}. This is particularly important for capturing the stochastic nature of atmospheric physics \citep{ma2019b, ma2022, ekanayaka2025multivariate} and characterizing extreme events. In this work, we view this problem as constructing a stochastic surrogate that maps input spatial fields (e.g., from GCMs) to target spatial fields (e.g., mimicking RCMs), thereby learning the conditional distribution encoded by the target source.

\paragraph{Limitations of existing approaches}

Current approaches to spatial downscaling and multi-scale data fusion are often too simplistic, insufficiently scalable, or data-hungry. 
Classical statistical emulators typically rely on Gaussian assumptions or linear dependence structures \citep[e.g.,][]{kennedy2000,le2014}. While Bayesian nonparametric treatments \citep{perdikaris2017} or hierarchical kriging approaches \citep{han2012} can offer more flexibility, they often struggle with scalability. For instance, methods based on cokriging or parallel partial cokriging \citep{ma2022} can be computationally infeasible for very large datasets, requiring recursive approximations \citep{cheng2023}. 
Furthermore, many of these models are designed for single training replicates (one observation per location), whereas paired ensemble data---which is increasingly common in Earth-system modeling---remains underutilized.
Conversely, recent deep-learning approaches \citep[e.g.,][]{niu2024, wu2022multi, buster2024} offer powerful nonlinear modeling capabilities but typically lack rigorous uncertainty quantification and require massive training data. In data-scarce regimes (e.g., limited RCM runs), these methods often require indiscriminate pooling over spatial regions or seasons, or sophisticated transfer learning. \citet{aich2024conditional} and \citet{mardani2024residual} propose diffusion-based generative approaches, but these can be computationally expensive and difficult to stabilize \citep{chattopadhyay2023long}.

\paragraph{Our method} 

We propose a generative multi-scale modeling approach via spatial autoregressive transport maps. We develop a probabilistic model that learns the non-Gaussian joint distribution and nonlinear dependence structure of nonstationary spatial fields across multiple scales. Our method requires only a small number of training samples by employing scale-aware autoregressive Gaussian processes (GPs) with carefully crafted regularization-inducing priors, which encourage data-dependent sparsity to achieve scalability.
Our approach builds upon recently proposed Bayesian triangular transport maps \citep{katzfuss2023}, which capture complex and non-Gaussian distributions of a single-resolution spatial field. We extend this to the multi-scale setting by sharing information hierarchically: we learn the conditional distribution at each scale given coarser-scale fields. While we use the terminology of ``coarse'' and ``fine'' scales for convenience, our framework is general: it links paired input and output fields, regardless of their relative spatial densities or whether they represent the same physical quantity. For instance, sparse observational data (e.g., from weather stations) could serve as the input to predict dense model outputs (e.g., reanalysis data), effectively reversing the traditional downscaling setup.
By using a conditional maximin ordering, which orders the input locations before the targets (e.g., ``coarse'' and ``fine''  scale locations, respectively, in the downscaling paradigm), we ensure an appropriate conditioning structure. This allows our method to serve as a highly efficient stochastic surrogate, capable of simulating realistic fine-scale climate behavior and quantifying uncertainty based on output from coarse global circulation models, substantially outperforming existing methods in numerical comparisons.

\paragraph{Outline of the paper}
The remainder of this document is organized as follows: Section \ref{sec:methodology} reviews Bayesian transport maps and introduces our model. Section \ref{sec:simulations} presents numerical comparisons on simulated datasets. Section \ref{sec:climate} evaluates model performances on multi-scale climate-model output. Section \ref{sec:conclusions} concludes with a discussion of future work. Implementation details of several competing models can be found in Appendix \ref{sec:app:competitors}. Code is available at (redacted for anonimization) \url{https://github.com/katzfuss-group/batram/tree/mf}

\section{Methodology}\label{sec:methodology}

\subsection{Scalable Bayesian transport maps: a review}\label{sec:review}

\paragraph{Motivation and core idea} 

Traditional Gaussian process (GP) models for spatial fields assume Gaussian marginal and joint distributions, which can be overly restrictive for many environmental and physical processes.
\citet{katzfuss2023} present a flexible probabilistic approach based on Bayesian transport maps that can learn non-Gaussian distributions and nonlinear dependence structures from limited training data, via estimation of a triangular transport map to a simple reference measure \citep[e.g.,][]{baptista2024representation,irons2022triangular,wang2022minimax}. The map is inferred by fitting a GP (auto-)regression at each spatial location. Carefully designed priors and Vecchia-type \citep{Vecchia1988} approximations allow the Bayesian transport map approach to scale to large spatial fields while fitting limited training data and quantifying uncertainty.

\paragraph{Maximin ordering}

Consider a (single-resolution) centered spatial field $\by = (y_1, \dots, y_N)^\top$ on the domain $\mathcal{D}$, where $y_i = y(\bs_i)$ for $\bs_i \in \mathcal{D}$.
\citet{katzfuss2023} propose to use a maximum-minimum-distance (maximin) ordering \citep{Guinness2016a} of $y_1, \ldots, y_N$ based on their spatial locations $\bs_1, \ldots, \bs_N$. This order can be interpreted as a coarse-to-fine ordering \citep[e.g.,][]{katzfuss2023,schafer2021}, where the scaling at each ordering index $i$ can be quantified by $\ell_i$, the distance of the $i$-th location to the nearest previously ordered neighbor. This core feature of the ordering will be used explicitly in our definition of the conditional maximin ordering in Section \ref{sec:mfargp}, where lower-resolution grids are assigned smaller length-scales. We assume that the entries of $\by$ are ordered according to the maximin ordering of $\bs_1, \ldots, \bs_N$. 

\paragraph{Bayesian specification and inference}
The Bayesian transport map (BTM) approach assumes a conditional Gaussian structure for each observation
\begin{equation*}
    y_i|\by_{<i}, f_i, d_i^2 \sim \normal(y_i|f_i(\by_{<i}), d_i^2)
\end{equation*}
and places conjugate GP-inverse-gamma priors on the mean functions $f_i: \mathbb{R}^{i-1}\to \mathbb{R}$ (with $f_1 \equiv 0$) and the standard deviations $d_i$:
\begin{equation*}
	\begin{aligned}
	\label{eq:model}
	f_i|d_i &\stackrel{ind}{\sim} \mathcal{GP}(0, d_i^2 K_i),\\
	d_i^2 &\stackrel{ind}{\sim} \mathcal{IG}(\alpha_i, \beta_i), \quad \alpha_i > 1, \beta_i > 0,
	\end{aligned}
\end{equation*}
for $i = 1, \dots, N$.
The inverse-gamma prior parameters $\alpha_i$ and $\beta_i$ encode a parametric decay in the $d_i^2$ induced by the maximin ordering.
The covariance kernels $K_i$ are designed to encode two principles:
(i) functions $f_i$ are pushed toward linearity as the minimum distance to previously ordered points decreases as $i$ increases;
(ii) input relevance of the $k$-th nearest (previously ordered) neighbor $y_{c_i(k)}$ decays exponentially with neighbor order $k$.
The latter property enables a (data-dependent) reduction in the conditioning sets $\by_{<i}$ encoding the screening effect \citep{stein2011} --- the diminishing relevance of distant neighbors in the presence of close neighbors. 

These design choices yield a parametrization of the model via six parameters, where two control the strength of non-linearities in the functions $f_i$, two control the decay rates of the conditional variances $d_i^2$, one controls the length-scale of the kernel, and one controls the conditioning set size. The specific form of the map components and the normal-inverse-gamma conjugacy enables factorization and closed-form evaluation of the marginal log-likelihood of the global parameter vector $\bftheta$ driving the inverse gamma prior on $d_i$ and the kernels $K_i$.
The factorization of the model evidence (i.e. the marginal log-likelihood) enables mini-batch training for scalable inference.
Alternatively, full Bayesian inference via MCMC can be carried out when accounting for uncertainty in $\bftheta$ is important.

\subsection{Multi-scale modeling via autoregressive transport maps}\label{sec:mfargp}

We extend the BTM framework from Section \ref{sec:review} to handle multi-scale spatial data by proposing scale-aware autoregressive GPs that can learn joint distributions across multiple resolutions or scales from limited training samples.

\paragraph{Multi-scale formulation}
Consider observations $\by = (\by_1^\top,\ldots,\by_R^\top)^\top$ across $R$ scales, resolutions, or data sources. These are ordered sequentially to reflect a continuum of availability and predictive effort: data that are easily accessible, computationally cheap, or spatially coarse appear first, while data that are increasingly difficult to obtain, expensive, or spatially fine appear last.  Under this hierarchical structure, any initially available sequence of, say, $r_0$ scales naturally serves as the input to predict the subsequent $R-r_0$ target scales. The $R$ scales correspond to sets of spatial locations $\mathcal{S}_1,\ldots,\mathcal{S}_R$. In other words, we have $\by_r = (y_{r,1},\ldots,y_{r,N_r})^\top$, where $y_{r,i}$ corresponds to spatial location $\bs_{r,i} \in \mathcal{S}_r \subset \mathcal{D}$.
We write the $N$-variate joint density, with $N = \sum_{r=1}^R N_r$, using an autoregressive decomposition as
\begin{equation*}%
	p(\by) = \prod_{r=1}^R p(\by_r | \by_{<r}),
\end{equation*}
where our goal is to learn these conditional distributions from a training ensemble $\{\by^{(j)}\}_{j=1}^n$ sampled independently from $p(\by)$, in the large-$N$, small-$n$ regime.
We make the typical Markov assumption $p(\by_r | \by_{<r}) = p(\by_r | \by_{r-1})$ across scales \citep{le2014, ma2022, perdikaris2017} for notational simplicity, but relaxing it is straightforward.

\paragraph{Relation to change-of-support}

Classical change-of-support (COS) models assume a know operator $\bA_r$ linking supports across resolutions (e.g., block averages), often enforcing $\by_{r-1} = \bA_r \by_{r}$, plus possibly noise \citep[e.g.,][]{gelfand2001,Gotway2002}. Our approach does not require specifying such an operator and does not enforce exact cross-resolution conservation. Instead, we learn cross-resolution dependence through the conditional model $p(\by_r|\by_{r-1})$ which can also capture settings where the mapping between resolutions may be unknown, approximate, or non-linear. When a COS operator is available and exact conservation is desired, incorporating $\bA$ as an structural constraint is a natural extension that would be interesting to explore in future work. 

\paragraph{Conditional maximin ordering}

\begin{figure}[htpb]
    \centering
    \includegraphics[width=0.8\linewidth]{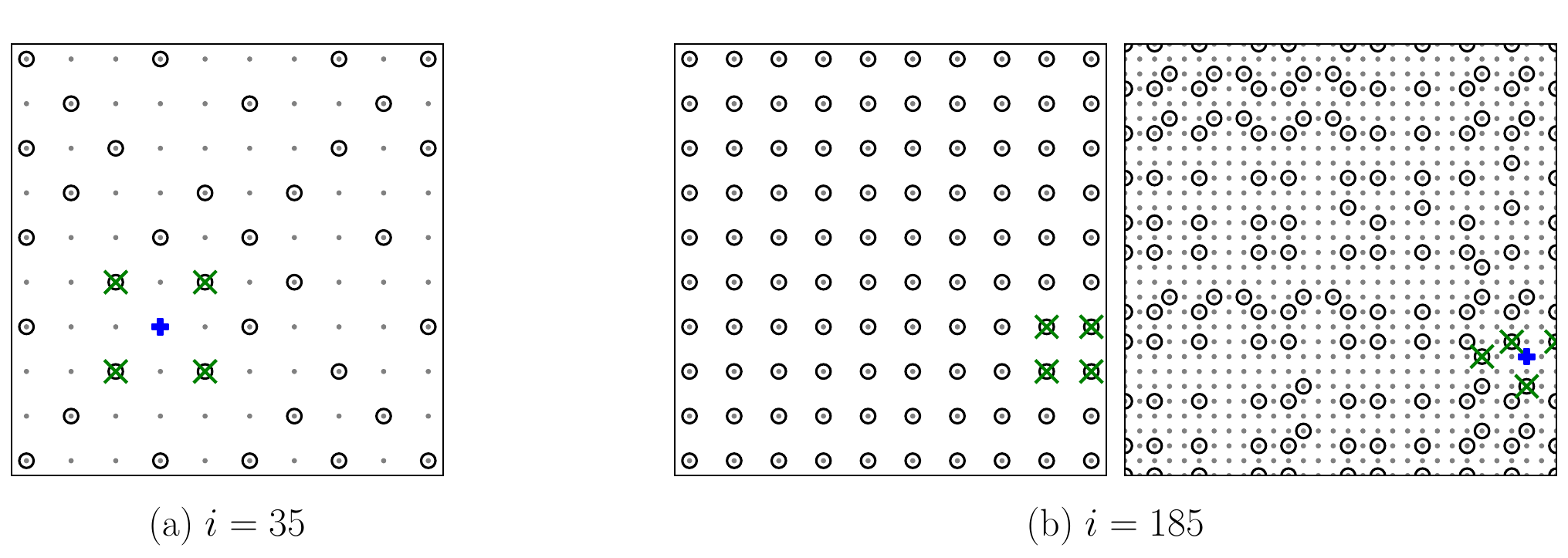}
    \caption{Conditional maximin ordering and conditioning sets for locations (gray dots) on a low-resolution grid of size $N_1 = 10 \times 10 = 100$ (left and middle panel) and a high-resolution grid (right panel) of size $N_2 = 30 \times 30 = 900$. For (a) $i=35$ and (b) $i=185$, the $i$th ordered location ($\color{blue}{+}$), the $i-1$ previously ordered locations ($\circ$), the $m_r=4$ nearest previously ordered locations $c_{ri}$ at the same scale ($\color{green}{\times}$ in the left and right panels), and the $m_r'=4$ nearest locations $c'_{ri}$ to $\bs_{185}$ in the lower scale $\locs_1$ ($\color{green}{\times}$ in the middle panel).}   
    \label{fig:maximin}
\end{figure}

We extend maximin ordering to the multi-scale setting through a conditional approach \citep{Schafer2020,chen2025sparse}, where points at lower-scales are ordered first, and conditional on those being ordered first, we sequentially order each subsequent resolution. The multi-scale maximin ordering is illustrated in Figure~\ref{fig:maximin}. The goal is to ensure that lower-resolution points anchor the ordering, so that every fine-scale location has nearby lower-resolution neighbors in its conditioning set. Formally, let $d(x, \mathcal{A}) := \min_{y \in \mathcal{A}} \|x-y\|$, and let $\mathcal{A}_{r-1} = \bigcup_{q < r} \mathcal{S}_q$, with $\mathcal{A}_1 = \emptyset$. For each scale, we produce the permutation $\pi_r: \{1, \dots, N_r \} \to \{1, \dots, N_r\}$ such that 
\begin{equation*}
	\begin{aligned}
	\pi_r(1) & = {\arg \max}_{i} d(\bs_{r,i}, \mathcal{A}_{r-1}), \\
    \pi_r(k+1) & = {\arg \max}_{i \notin \pi_r(\{1:k \} )} d(\bs_{r,i}, \mathcal{A}_{r-1} \cup \{\bs_{r,\pi_r(1)}, \dots, \bs_{r,\pi_r(k)}\}), \quad k=1,\ldots,N_r-1.
	\end{aligned}
\end{equation*}
For $\pi_1(1)$ we pick an arbitrary location, usually near the center of the lowest-scale grid. The conditional maximin ordering is then given by the global permutation $P: \{1, \dots, N\} \to \bigcup_{r=1}^R \{ (r, i) \}$, where 
\begin{equation*}
    P(\sum_{q=1}^r N_q + k  ) = (r, \pi_r(k)), \quad 1 \leq r \leq R,  \: 1 \leq k \leq N_r.
\end{equation*}
Hereafter, $\by$ and locations $\mathcal{S}$ are assumed to be ordered according to this conditional maximin formulation. We define the lengthscales (i.e., distance to the nearest previously ordered point) as
\begin{equation*}
    \ell_{r,i} = d(\bs_{r,i}, \mathcal{A}_{r-1} \cup \{\bs_{r,j}: j<i \}).
\end{equation*}
The conditional maximin ordering can be constructed in $\mathcal{O}(N \log^2(N) )$ time and $\mathcal{O}(N)$ space \citep{chen2025sparse,Schafer2020}. The ordering is precomputed once before training, and hence it is usually not a computational bottleneck. 

\paragraph{Multi-scale transport map}

We propose to learn the multi-scale distribution $p(\by)$ via a transport map $\mathcal{T}: \mathbb{R}^N \to \mathbb{R}^N$ to the standard multivariate normal distribution: $\mathcal{T}(\by) \sim \normal_N(\bfzero,\bI_N)$, in an extension of  \citep{katzfuss2023}. Without loss of generality, we consider a lower-triangular map
\begin{equation}\label{eq:tm}
    \mathcal{T}(\by) = \begin{bmatrix*}[l]
        \mathcal{T}_{1,1}(y_{1,1}) \\
        \quad \vdots  \\
        \mathcal{T}_{1,N_1}(y_{1,1}, \dots, y_{1,N_1}) \\
        \mathcal{T}_{2,1}(y_{1,1}, \dots, y_{1,N_1}, y_{2,1}) \\
        \quad\vdots \\
        \mathcal{T}_{R,N_R}(y_{1,1}, \dots, y_{1,N_1}, y_{2,1}, \dots, y_{2,N_2}, \dots, y_{R,1}, \dots, y_{R,N_R})
    \end{bmatrix*}.
\end{equation}
To ensure invertibility of the map, we define the monotonic (in the last argument) map components
\begin{equation}\label{eq:mf-tm-component}
\mathcal{T}_{r,i}(y_{1,1}, \ldots, y_{r,i}) = \frac{y_{r,i} - f_{r,i}(\by_{\tilde{c}_{r,i}})}{d_{r,i}},
\end{equation}
where the conditioning set $\by_{\tilde{c}_{r,i}} = (\by_{r,c_{r,i}}, \by_{r-1,c'_{ri}})$ contains the $m_r$ nearest previously ordered neighbors of the same resolution and $m'_r$ nearest neighbors of the previous resolution to location $\bs_{r,i}$, with $c_{r,i}$ and $c'_{r,i}$ denoting the corresponding index sets; hence we have $f_{r,i}: \mathbb{R}^{\min\{m_r, i \}+m_r'} \to \mathbb{R}$ with $f_{1,1} \equiv 0$.
This yields a factorization where each element in $\by$ can be expressed by a normal distribution when conditioning on values $\by_{\tilde{c}_{r,i}}$ as well as the relevant function $f_{r,i}$ and variance term $d^2_{r,i}$,
\begin{equation*}%
y_{r,i}|\by_{\tilde{c}_{r,i}}, f_{r,i}, d^2_{r,i} \sim \mathcal{N}(y_{r,i} | f_{r,i}(\by_{\tilde{c}_{r,i}}), d_{r,i}^2).
\end{equation*}

\begin{figure}[htpb]
    \centering
    \includegraphics[width=.64\linewidth]{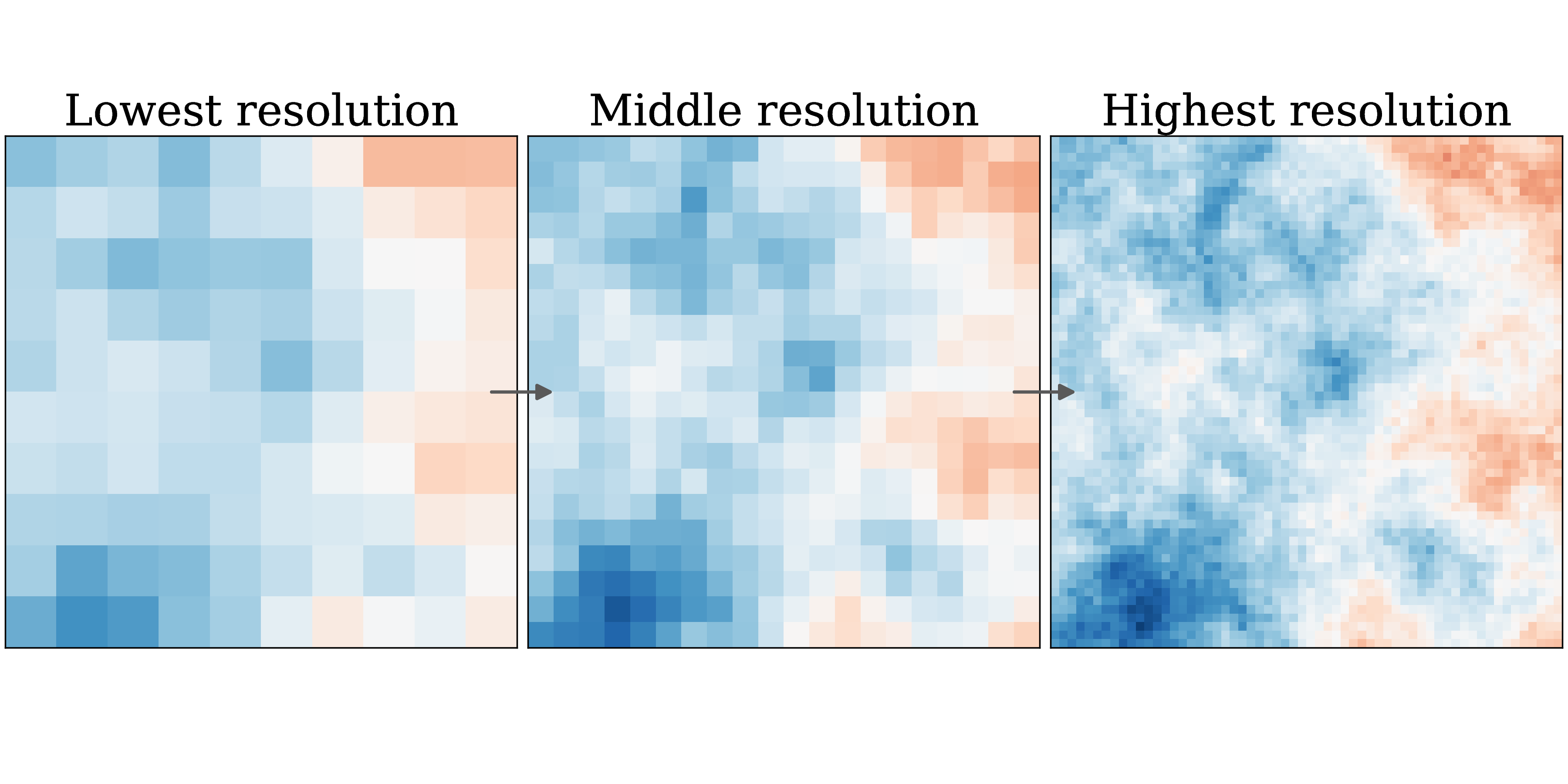}
    \includegraphics[width=.74\linewidth]{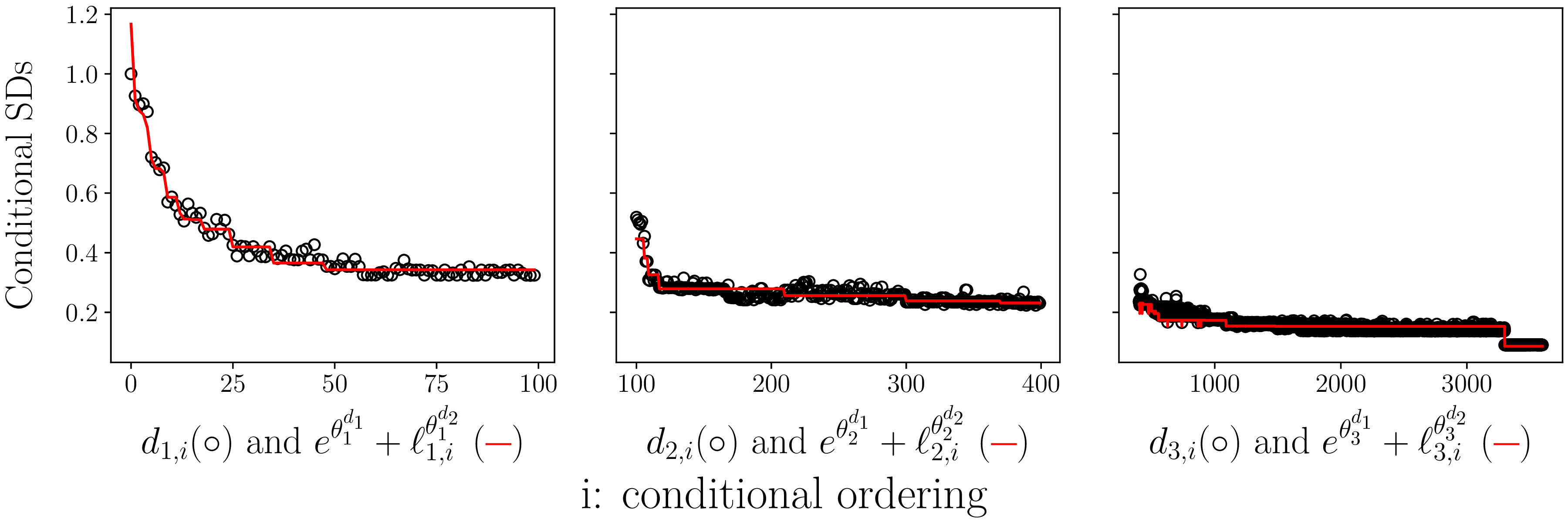}
    \includegraphics[width=.74\linewidth]{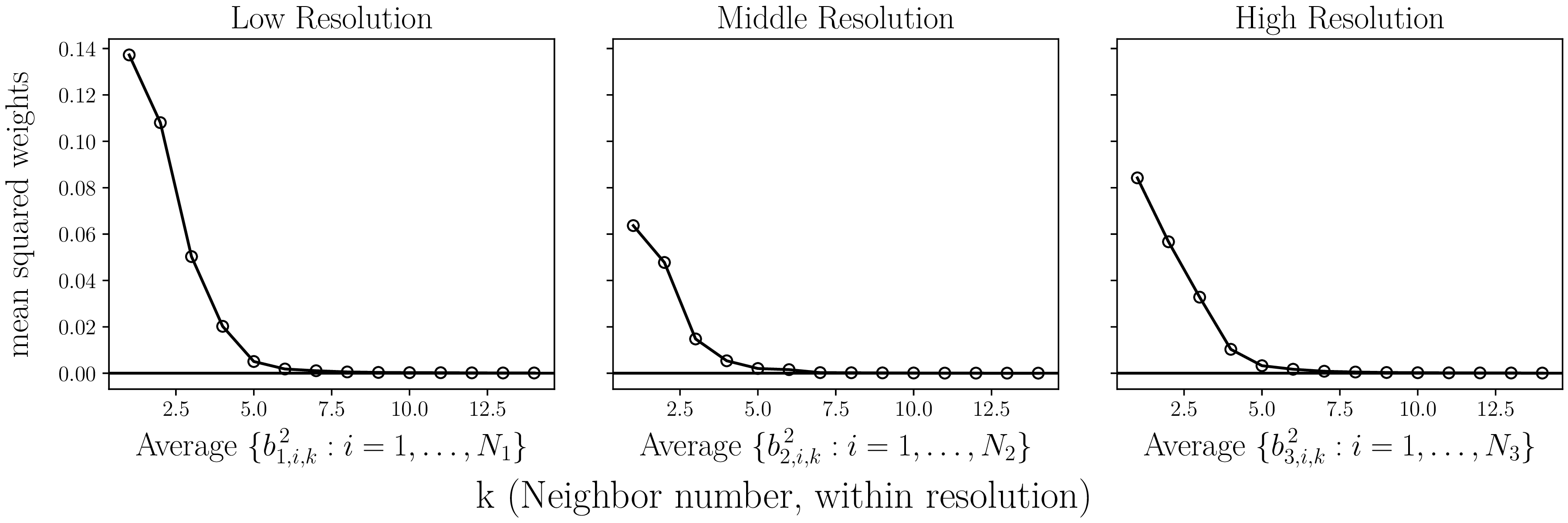}
    \includegraphics[width=.54\linewidth]{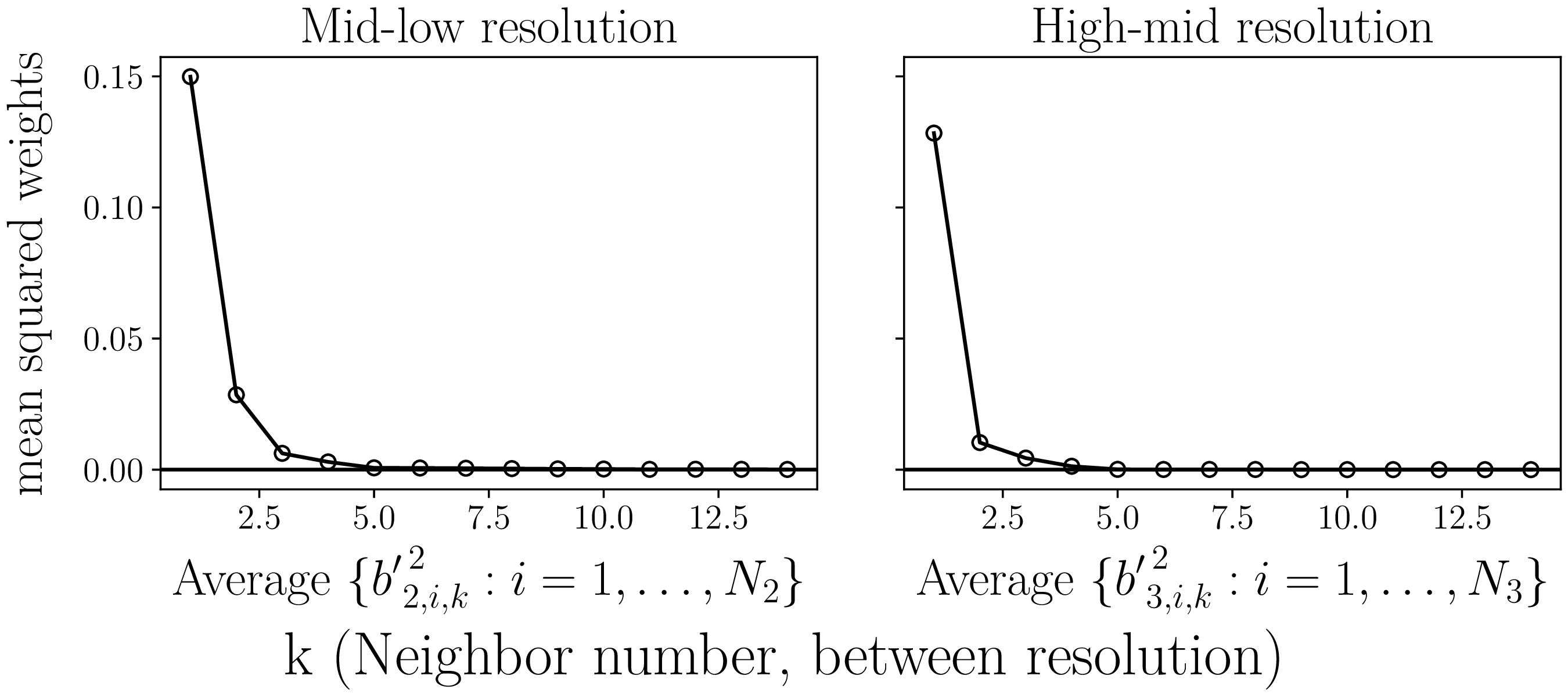}
    \caption[]{For a GP with exponential covariance on coarser-to-finer grids and with the maximin ordering in Figure \ref{fig:maximin}, the map components \eqref{eq:mf-tm-component} can be written as
    \begin{minipage}{\linewidth}
    \begin{equation*}
        f_{r,i}(\by_{r, <i}, \by_{r-1}) = \sum_{k=1}^{i-1} y_{r,c_{r,i}(k)} b_{r,i,k} + \sum_{k=1}^{N_{r-1}} y_{r-1,c^\prime_{r,i}(k)} b^\prime_{r,i,k}.
    \end{equation*}
    \end{minipage}
     For the $i$th location in the $r$th-scale, $c_{r,i}(k)$ indicates the $k$th previously ordered nearest neighbor in the same resolution, and $b_{r,i,k}$ the corresponding kriging weight; while $c^\prime_{r,i}(k)$ indicates the $k$-th nearest neighbor in the previous resolution, with $b^\prime_{r,i,k}$ as the corresponding kriging weight.
    (First row): Sample from the process.
    (Second row): The conditional standard deviations $d_{r,i}$ decay polynomially as a function of $\ell_{r,i}$. (The sudden drop around $i=3{,}200$ is due to grids at different resolutions sharing locations.)
    (Third row): Squared kriging weights in the same resolution decay rapidly in neighbor number. 
    (Fourth row): Squared kriging weights in previous resolution also decay rapidly with neighbor number.}
    \label{fig:assumption}
\end{figure}

Using these small conditioning sets is motivated by the so-called screening effect \citep{stein2002,stein2011}, where the relevance of distant neighbors decays when conditioning on closer neighbors (see Figure \ref{fig:assumption}). Under the conditional maximin ordering, the rates at which these relevances decay are exponential in certain settings, as shown in \citet{schafer2021} and \citet{chen2025sparse}, but with constants depending on the specific data-generating process. Similarly, \citet{chen2024precision} show that methods similar to our transport map approach only need a polylogarithmic number of samples (in $N$) to accurately estimate the distribution in the Gaussian setting.   

\paragraph{Scale-aware priors}
We specify independent conjugate priors for each scale, 
\begin{equation*}
\begin{aligned}
	f_{r,i}|d_{r,i} &\sim \mathcal{GP}(0, d_{r,i}^2 K_{r,i}), \label{eq:mf-prior}\\
	d_{r,i}^2 &\sim \mathcal{IG}(\alpha_{r,i}, \beta_{r,i}),
\end{aligned}
\end{equation*}
whose specific parameterizations are motivated by the behavior of aggregated Mat\'ern Gaussian fields (as illustrated in Figure \ref{fig:assumption}), with parameters that are shared within resolutions but differ across resolutions to capture scale-dependent behavior.

Following the decay principles from the BTM reviewed in Section~\ref{sec:review}, we parameterize the inverse-gamma parameters using the minimum neighbor distance $\ell_{r,i}$ by
\begin{equation*}%
\mathbb{E}[d_{r,i}^2] = e^{\theta^{d_1}_{r}} \ell_{r,i}^{\theta^{d_2}_{r}},
\end{equation*}
where the parameters $\theta^{d_1}_{r}$ and $\theta^{d_2}_{r} > 0$ control scale-specific decay rates of the residual variance. Note that we use superscripts on $\theta$s to denote parameter names, not exponents. This polynomial decay specification is motivated by theoretical results showing that the large class of stochastic processes with quasi-quadratic log-likelihood functions exhibit polynomial decay of conditional variances under the conditional maximin ordering \citep[][Lemmas 5.10 and 5.15]{schafer2021}, where GPs with covariance functions given by the Green's function of elliptic PDEs of order $s$ have conditional variances of order $\ell_{r,i}^{2sr}$ measured at scale $r$.

Empirical support for this decay is illustrated in Figure \ref{fig:assumption}. To parameterize the gamma prior for $d_{r,i}^2$, we assume that the prior standard deviation equals $g$ times the prior mean, resulting in $\alpha_{r,i} = 2 + 1/g^2$ and $\beta_{r,i} = \exp(\theta^{d_1}_{r}) \ell_{r,i}^{\theta^{d_2}_{r}}(1 + 1/g^2)$. Following \citet{katzfuss2023}, we let $g = 4$ to obtain a relatively vague prior for the conditional variances. 

The covariance kernels $K_{ri}$ extend the $K_i$ from \citet{katzfuss2023} to handle inputs from multiple resolutions and by introducing scale-specific parameters, resulting in
\begin{equation}\label{eq:mf-kernel}
	K_{r,i}(\by_{\tilde{c}_{r,i}}, \by'_{\tilde{c}_{r,i}}) =
		\by_{\tilde{c}_{r,i}}^\top \bQ_r \by'_{\tilde{c}_{r,i}} +
		\sigma_{r,i}^2 \rho\left(
		\sqrt{(\by_{\tilde{c}_{r,i}} - \by'_{\tilde{c}_{r,i}})^\top \bQ_r (\by_{\tilde{c}_{r,i}} - \by'_{\tilde{c}_{r,i}})} / \gamma_r
		\right),
\end{equation}
where $\rho$ is an isotropic correlation function with resolution-specific range parameter $\gamma_r = \exp(\theta_{r}^{\gamma})$. (In our numerical experiments, we take $\rho$ to be the Matérn correlation with smoothness $\nu = 3/2$.)
The nonlinearity variance $\sigma_{r,i}^2 = e^{\theta^{\sigma_1}_r} \ell_{r,i}^{\theta^{\sigma_2}_r}$ in the kernels decays when $r$ and $i$ increase, pushing functions toward linearity at fine scales. This behavior is justified from the fact that, for many physical processes, if we condition on averages over subdomains of decreasing diameter, we observe that the conditional distributions quickly become Gaussian, as argued by \citet{katzfuss2023} as a consequence of the Poincaré inequality \citep{adams2003sobolev}. These averages over subdomains can be viewed as observations in lower resolutions or earlier-order points in the conditional maximin ordering in the current scale.

We propose the relevance matrix $\bQ_{r}$ as the diagonal matrix with weights $q_{r,j}$, $j=1,\dots, m_r + m'_r$, decaying exponentially with neighbor number following the screening effect with respect to each scale,
$$
q_{r,j} = \begin{cases}
	\exp(\theta^{q_0}_{r} - \theta^{q_1}_{r} j),         & j \leq m_r,\\
	\exp(\theta^{q'_0}_{r} - \theta^{q'_1}_{r} (j - m_r)), & m_r < j \leq m_r + m'_r.
\end{cases}
$$
The conditioning set sizes are determined adaptively as $m_r = \max\{k : \exp(\theta_{r}^{q_0} - \theta_{r}^{q_1} k) \geq \varepsilon\}$
and similarly for $m'_r$, 
where $\varepsilon \geq 0$ is specified upfront.
In our numerical experiments, we set $\varepsilon = 0.01$, which typically results in conditioning sets of size less than 20 while maintaining high inferential accuracy. The intercept terms $\theta_r^{q_0}, \theta_r^{q_0^\prime}$ allow extra flexibility for the model, as it is a data dependent way to pick scale-specific truncation radii.

We motivate regularization and sparsity by exploiting the so-called screening effect \citep[e.g.,][]{stein2002,stein2011}. 
For multi-resolution spatial observations, \citet[][Example 5.1]{schafer2021} establish exponential screening rates for processes derived from elliptic boundary-value problems observed at different scales, induced by the conditional maximin ordering.
Under the assumption that observation points are approximately homogeneously distributed within each scale, we expect similar conditional independence properties to hold at fine scales.
Importantly, the arguments presented by \citet{schafer2021} specifically accommodate settings where the coarser-scale observations represent local averages rather than point-wise evaluations of the underlying function, which matches the empirical scenarios examined in Section~\ref{sec:coarsening} and illustrated in Figure~\ref{fig:assumption}. 

\paragraph{Scalable inference}

The multi-scale model requires $9R-2$ parameters total (9 per resolution except for the first, which has no cross-resolution terms), which are estimated from training data $\by^{(1:n)} = \{\by^{(j)}\}_{j=1}^n$. The normal-inverse-gamma conjugacy enables closed-form marginal likelihood evaluation, allowing empirical Bayes inference via gradient-based optimization of $\log p(\by^{(1:n)} | \bftheta)$, where it can be shown by extending results in \citet{katzfuss2023} that
\begin{equation}
p(\by^{(1:n)} | \bftheta) = \prod_{r=1}^{R}\prod_{i=1}^{N_r} (|\bG_{r,i}|^{-1/2} \times (\beta_{r,i}^{\alpha_{r,i}} / \tilde{\beta}_{r,i}^{\alpha_{r,i}}) \Gamma(\tilde{\alpha_{r,i}}) / \Gamma(\alpha_{r,i})),\label{eq:integratedlikelihood}
\end{equation}
with $\bG_{r,i} = \bK_{r,i} + \bI_n$, $\bK_{r,i} = \big(K_{r,i}(\by_{ \tilde{c}_{r,i}}^{(j)}, \by_{ \tilde{c}_{r,i}}^{(l)})\big)_{j,l = 1, \dots,  n}$, $\tilde{\alpha}_{r,i} = \alpha_{r,i} + n/2$, and $\tilde{\beta}_{r,i} = \beta_{r,i} + \by_{r,i}^\top \bG_{r,i}^{-1} \by_{r,i}/2$. 
We write $\bftheta$ for the vector of all parameters, and $\bftheta_r$ for the parameters relevant for the $r$-th resolution. 
Because $\bftheta_r$ and $\bftheta_s$ for $r \neq s$ do not appear in the same terms in \eqref{eq:integratedlikelihood}, training across scales can proceed independently and can be parallelized with computational cost $\mathcal{O}(N_r(n^3 + \tilde{m}_r n^2))$ per scale, where $\tilde{m}_r = m_r + m'_r$ typically remains below 30.
Within each resolution, mini-batch optimization provides additional speedups, reducing the computational cost to $\mathcal{O}(B(n^3 + \tilde{m}_r n^2))$ per gradient update for batch size $B$, containing only information at $B$ random spatial locations for fidelity $r$. Since higher resolutions typically contain vastly more spatial locations than lower ones, the most substantial computational gains can be achieved at the finer scales, which constitute the primary computational bottleneck. Following \citet{goyal2017accurate}, we scale the learning rate linearly with the batch size. 

After training (i.e., obtaining the empirical Bayes estimator $\hat{\bftheta}$), the posterior predictive distribution
\begin{equation*}%
p(\by|\by^{(1:n)}, \hat\bftheta) = \prod_{r=1}^R p(\by_{r}|\by_{r-1}, \by^{(1:n)}, \hat\bftheta) = \prod_{r=1}^R \prod_{i=1}^{N_r} p(y_{r,i}|\by_{\tilde{c}_{r,i}}, \by^{(1:n)}, \hat\bftheta)
\end{equation*}
can be used for efficient marginal likelihood evaluation at test samples, posterior sampling, and conditional inference. Individual product terms can be obtained by straightforward extension of \citet[Prop.~1]{katzfuss2023}.
Importantly, one can readily compute the distribution of higher-scale observations conditional on lower-resolution data. When modeling uncertainty quantification is critical, the framework can be extended using MCMC or Laplace approximation to account for parameter uncertainty.

\section{Simulation study}\label{sec:simulations}

\subsection{Comparison metrics and methods}\label{sec:metrics}

We compare several different estimators $\hat{p}$ of true data-generating distributions $p(\by)$ and associated conditionals $p(\by_r|\by_{<r})$. To evaluate the performance of an estimator, we compute the average negative log-density of the learned distribution on a set of held-out test samples; for one test sample $\by^*$, we have $- \log \hat{p}(\by^*) = -\sum_{r=1}^R \log \hat{p}(\by_r^*|\by_{<r}^*)$. This is a strictly proper scoring rule for probabilistic prediction, called the log-score \citep[e.g.,][]{Gneiting2014}, and provides an approximation, up to an additive constant, of the Kullback-Leibler (KL) divergence $\rm{KL}(p\|\hat{p})$. This allows us to compare the accuracy and goodness-of-fit of the learned distribution $\hat{p}$ against the true data-generating distribution $p$. While conditional log-scores $\log \hat{p}(\by_r^*|\by_{<r}^*)$ can be evaluated separately or cost-weighted, the highest-resolution field empirically dominates the total log-score in our experiments; moreover, our method consistently outperforms competitors across all resolutions $r$. The uncertainty in the log scores is negligible relative to the differences between the log scores, so we do not report it. We also plot generated conditional spatial fields from our method for a qualitative assessment of its accuracy. 

The methods to be compared (see Appendix \ref{sec:app:competitors} for implementation details) are: 
\begin{description}
    \item[MS BTM:] Our multi-scale Bayesian transport map method, as described in Section \ref{sec:mfargp}.
    \item[HK:] Our implementation of hierarchical kriging \citep{han2012}.
    \item[NARGP:] Our implementation of the nonlinear autoregressive GP  \citep{perdikaris2017}
    \item[VAE:] A multi-scale variational auto-encoder, inspired by \citet{cheng2024bi}.
    \item[Matérn:] Fits a GP with Matérn covariance to the highest-resolution observations and assumes the lower-resolutions are block-wise averages of the higher-resolutions, similar to \citep{ma2019b}. Not shown for all comparisons because of poor performance when this assumption is not met. 
\end{description}

\subsection{Downscaling of block averages}\label{sec:coarsening}

In many downscaling and multi-resolution problems in the geosciences, the average of the fine-scale values in a given cell is equal to the corresponding value at a coarser scale. With this in mind, we consider three scales such that we observe the highest resolution on a regular grid of size $N_3 = 30 \times 30 = 900$, the middle resolution on a grid of size $N_2 = 10 \times 10 = 100$, and the lowest resolution on a grid of $N_1 = 5 \times 5 = 25$. All grids are on the unit square. For the highest-scale data $\by_3$, the data-generating process is a simulation scenario called NR900 in \citet{katzfuss2023}, where each conditional mean function $f_{3i}$ in \eqref{eq:mf-tm-component} is specified as 
$
f_{3,i}(\by_{3,<i}) = \bb_{3,i}^\top \by_{c_{3,i}} + 2 \sin(4(b_{3,i,1} y_{c_{3,i}(1)} + b_{3,i,2} y_{c_{3,i}(2)})),
$
and $\bb_{3,i} = (b_{3,i,1}, \dots, b_{3,i,|c_{3,i}|})^\top$ is obtained from a GP with exponential covariance with range $0.3$. Given a $\by_3$ generated in this way, we take averages within 100 spatial blocks of size $3 \times 3$ to obtain $\by_2$, and then averages within 25 blocks of size $2\times 2$ to obtain $\by_1$. 

\begin{figure}[htpb]
    \centering
    \includegraphics[width=0.68\linewidth]{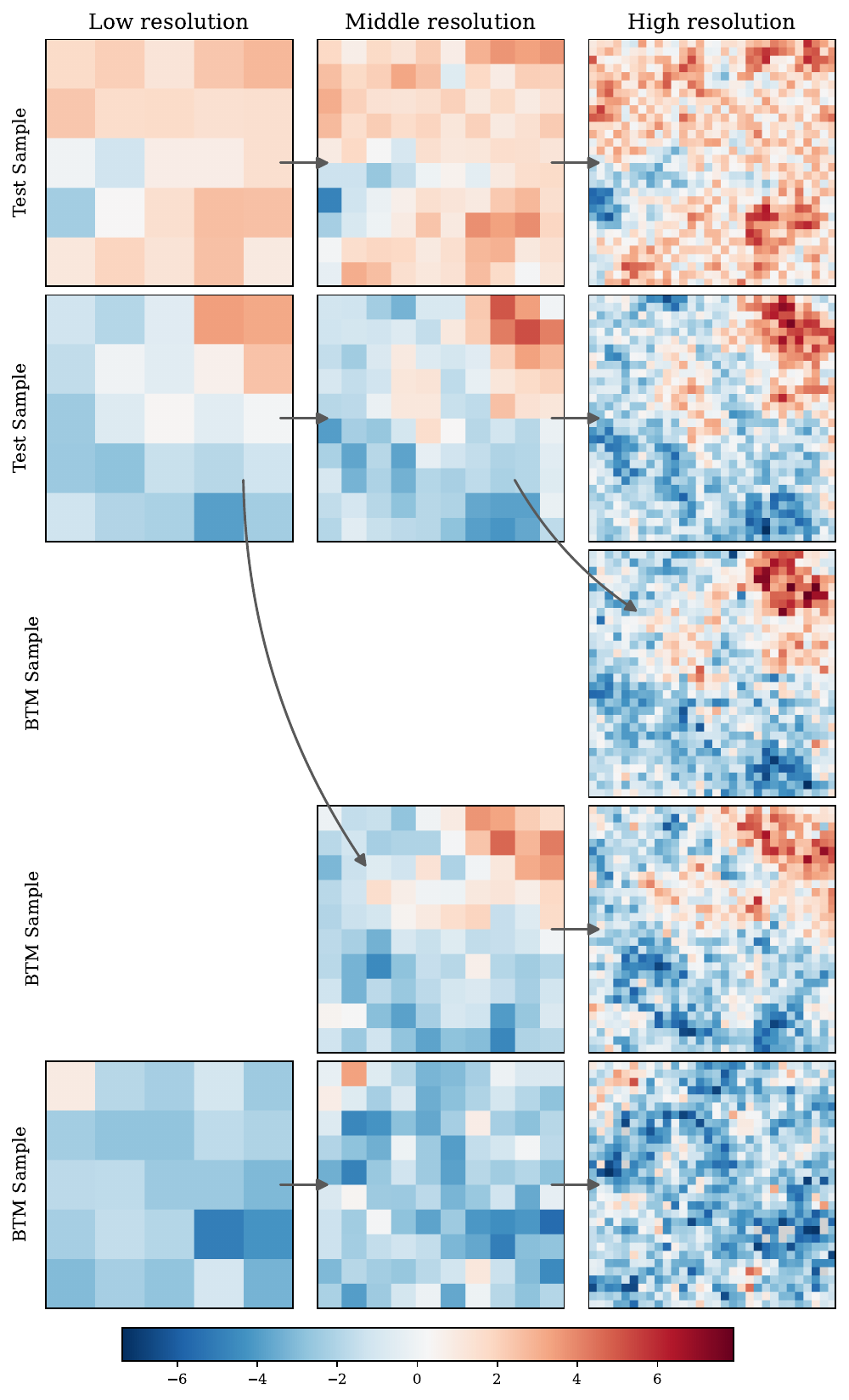}
    \caption{\textbf{For downscaling of (linear) block averages, samples from our BTM mimic those from the true data-generating process, and they get closer to test samples when conditioning on more lower-scale test fields.} The first two rows show test samples from the block-averaging data-generating process described in Section \ref{sec:coarsening}. The next three rows are samples from our model (trained on $50$ samples). Arrows represent conditioning relationships (e.g., the high-resolution BTM sample in row 3 is drawn from the conditional distribution given the middle-resolution test sample in row 2). More concretely, comparing column 3 in rows 2, 3, 4, rows 3 and 4 are able to reproduce the global and local features in row 2 (e.g., high values in the top right), with row 3 matching the ``true'' sample in row 2 more closely, as it conditions on more information than row 4.}
    \label{fig:samples_BTM}
\end{figure}

Figure \ref{fig:samples_BTM} shows that the samples generated from the trained model are qualitatively similar to the observed samples. As expected, samples become qualitatively closer to the test fields when conditioning on low-resolution data, because we fix some of the first map coefficients \citep[e.g.,][]{Marzouk2016,baptista2020conditional}, which amounts to fixing the large-scale features of an observed field. Notably, the sampled fields approximately respect the structure of the generating process, where lower scales are averages of cells in higher scales, even though this relationship is not directly encoded in the model. 

\begin{figure}[htpb]
    \centering
    \includegraphics[width=\linewidth]{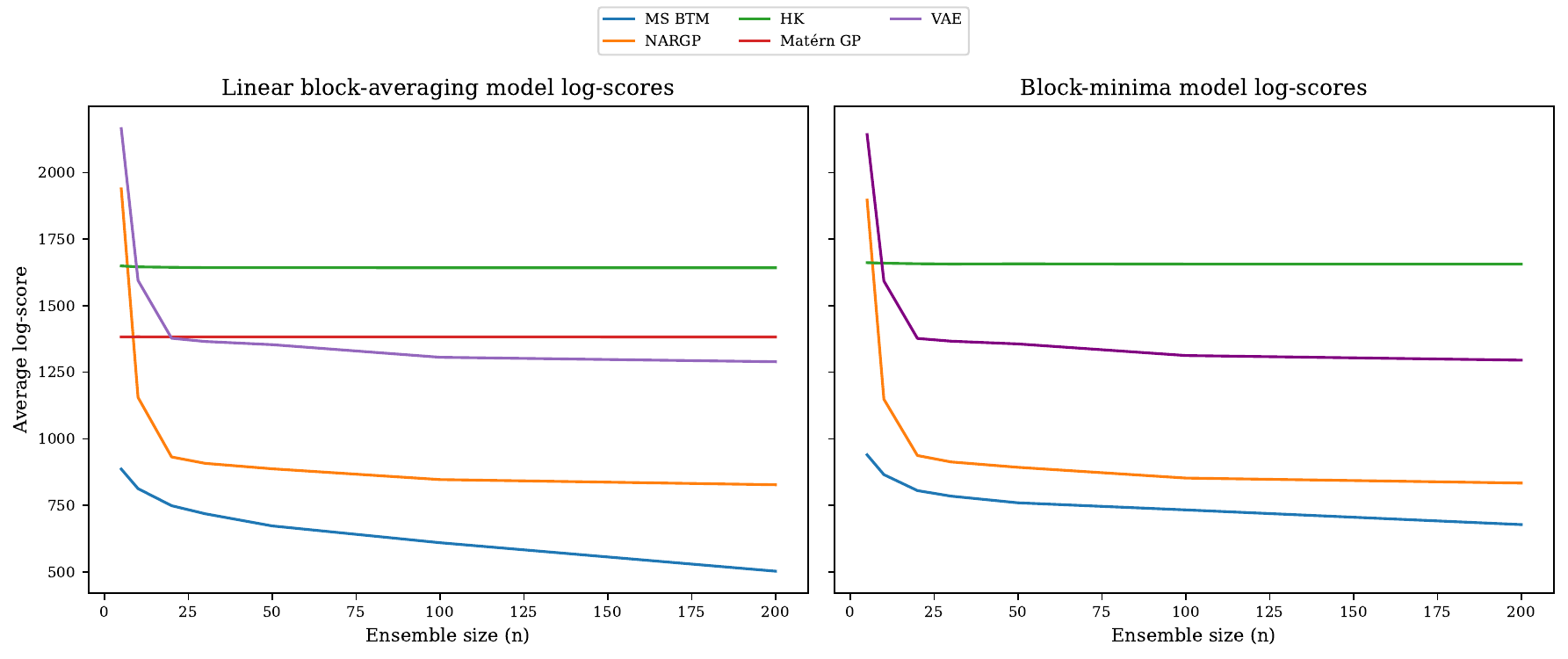}
    \caption{\textbf{Our multi-scale Bayesian transport map (MS BTM) performs best in terms of log-scores for all sample sizes $n$ in two simulation scenarios.} (Left): Block-averaging scenario from Section~\ref{sec:coarsening}. (Right): (Nonlinear) block minima from Section~\ref{sec:min}. The competing methods are listed in Section~\ref{sec:metrics}.}
    \label{fig:comparison}
\end{figure}

We compare the accuracy of the methods described in Section~\ref{sec:metrics} for different training  ensemble sizes $n$, computing the average log-score of the model on 50 test fields.
See the left part of Figure \ref{fig:comparison} for the comparison between models, where our multi-scale transport map is the best-performing model for all ensemble sizes, with NARGP model being the second best for big $n$, and the Matérn model (which is the only model that a priori encodes the true averaging relationship between resolutions) is the second best for small $n$. The VAE is not competitive with NARGP or MS BTM, likely because deep models typically require a large number of training samples to achieve substantial predictive power, which is not feasible in our data-scarce regime. 

\subsection{Downscaling of (nonlinear) block minima }\label{sec:min}

Our model can learn non-linear relationships between scales, via the nonlinear term in the kernel in \eqref{eq:mf-kernel}, where some of the neighbors in $\tilde{c}_{r,i}$ correspond to nearest neighbors in the $r-1$-th resolution for the $r$-th kernel. 

For this scenario, we use the same high-resolution simulations described in Section \ref{sec:coarsening}, but instead of averaging the values in subgrids of the highest-resolution, we coarsen by using the minima of these subgrids. 

\begin{figure}[htpb]
    \centering
    \includegraphics[width=0.73\linewidth]{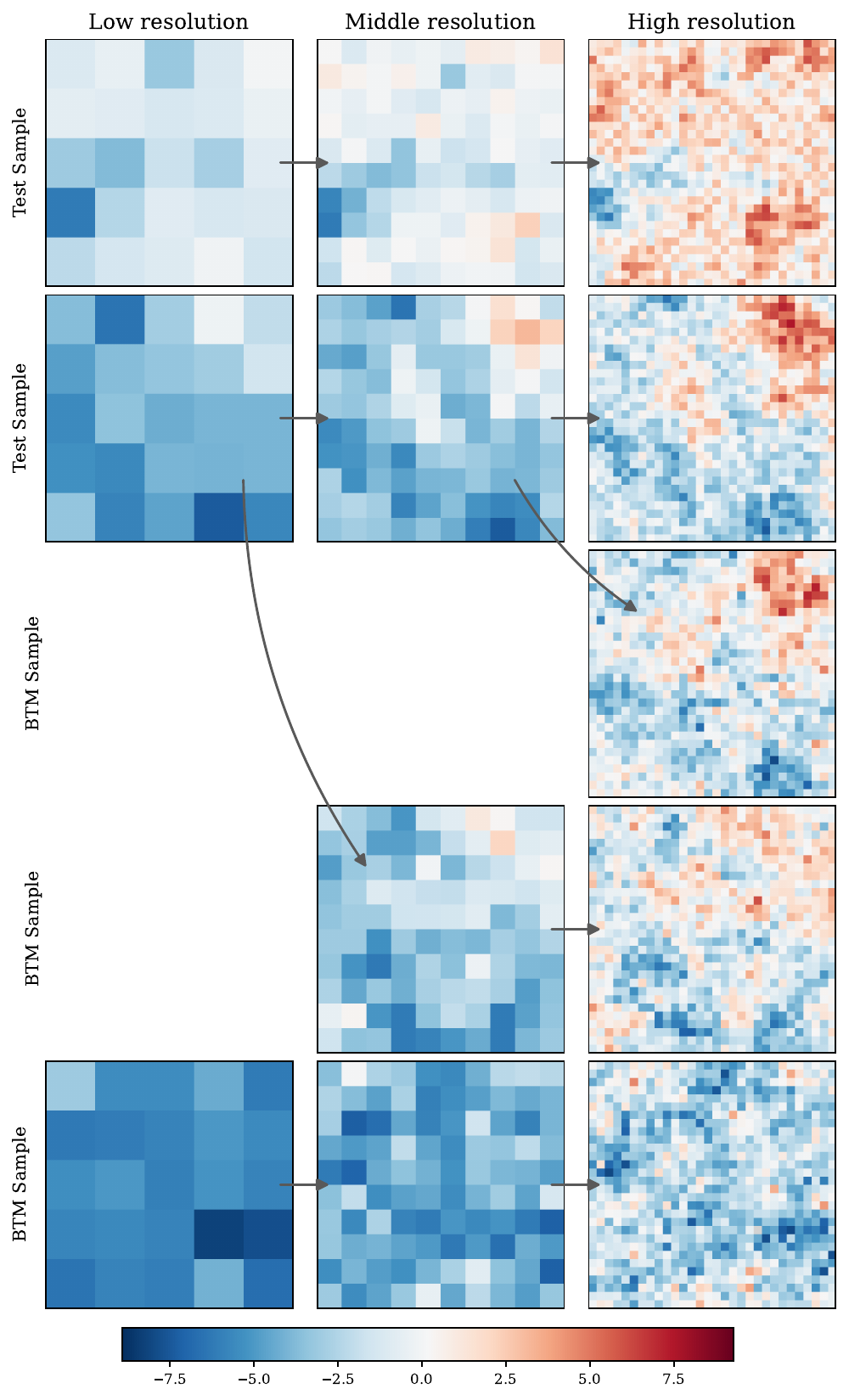}
    \caption{\textbf{For multi-scale conditionals of (nonlinear) block minima, samples from our BTM also mimic those from the true data-generating process.}
    The first two rows show test samples from the block-minima data-generating process described in Section \ref{sec:min}. The next three rows are samples from our model (trained on $50$ samples). Arrows represent conditioning relationships. Comparing to Figure \ref{fig:samples_BTM}, it is more difficult for the MS BTM to reconstruct local features (e.g., the high values in the top right in row 2, column 3), but it successfully reproduces some of the highly nonlinear structure. }
    \label{fig:samples_min}
\end{figure}

Figure \ref{fig:samples_min} shows that the generated samples are qualitatively similar to the observed samples. Our approach successfully reconstructs the higher-resolution samples, because it is able to model the nonlinear structure between scales. The right panel of Figure \ref{fig:comparison} confirms that the multi-scale BTM outperforms all competing models for each training ensemble size in this nonlinear scenario. 

\section{Stochastic surrogates for regional climate models}\label{sec:climate}

Climate models are essential computational tools for simulating and studying Earth's climate and are used for climate research and policy-making. They are, in essence, computer code that models Earth-system behavior via differential equations. Producing ensembles of samples using such models is very computationally costly \citep[e.g.,][]{kay2015community}, but they are needed to assess uncertainty related to different initial conditions. Stochastic surrogates, such as the one presented in this work, allow us to summarize, study, and replicate the distribution to generate additional samples at much lower computational cost. 

However, as noted by \cite{yuUpperTailPrecipitation2020}, to capture some quantities of interest, especially those with severe impacts, fine-scaled grid spacing is necessary, usually accomplished via dynamical downscaling, where GCM output drive or force high-resolution RCM \citep{xu2019dynamical}. However, this increase in resolution in climate models incurs a substantial computational cost, and a $\times 10$ decrease in grid spacing can lead to a $\times 1000$ increase in computational cost \citep{schneider2024opinion}. 

We interpret GCMs and RCMs as low-scale and high-scale encodings, respectively, of Earth's climate distribution, and an RCM run forced by GCM output as a sample from the conditional distribution of the high-resolution field given the coarse low-resolution GCM field. Our method can be used as a stochastic surrogate that produces RCM-like samples given GCM realizations. 

We use the 50 GCM members of the CanESM2 \citep{chylek2011observed} Large Ensemble (CanESM2-LE) \citep{kushner2018canadian,kirchmeier2017attribution} produced by the Canadian Centre for Climate Modeling and Analysis under the RCP8.5 emission scenario. \cite{leduc2019climex} uses the CanESM2-LE output to drive the Canadian Regional Climate Model (CRMC5) \citep{martynov2013reanalysis} and obtain the CRCM5 Large Ensemble (CRMC5-LE), which we use as our RCM data. 

\begin{figure}[htpb]
    \centering
    \includegraphics[width=0.85\textwidth]  {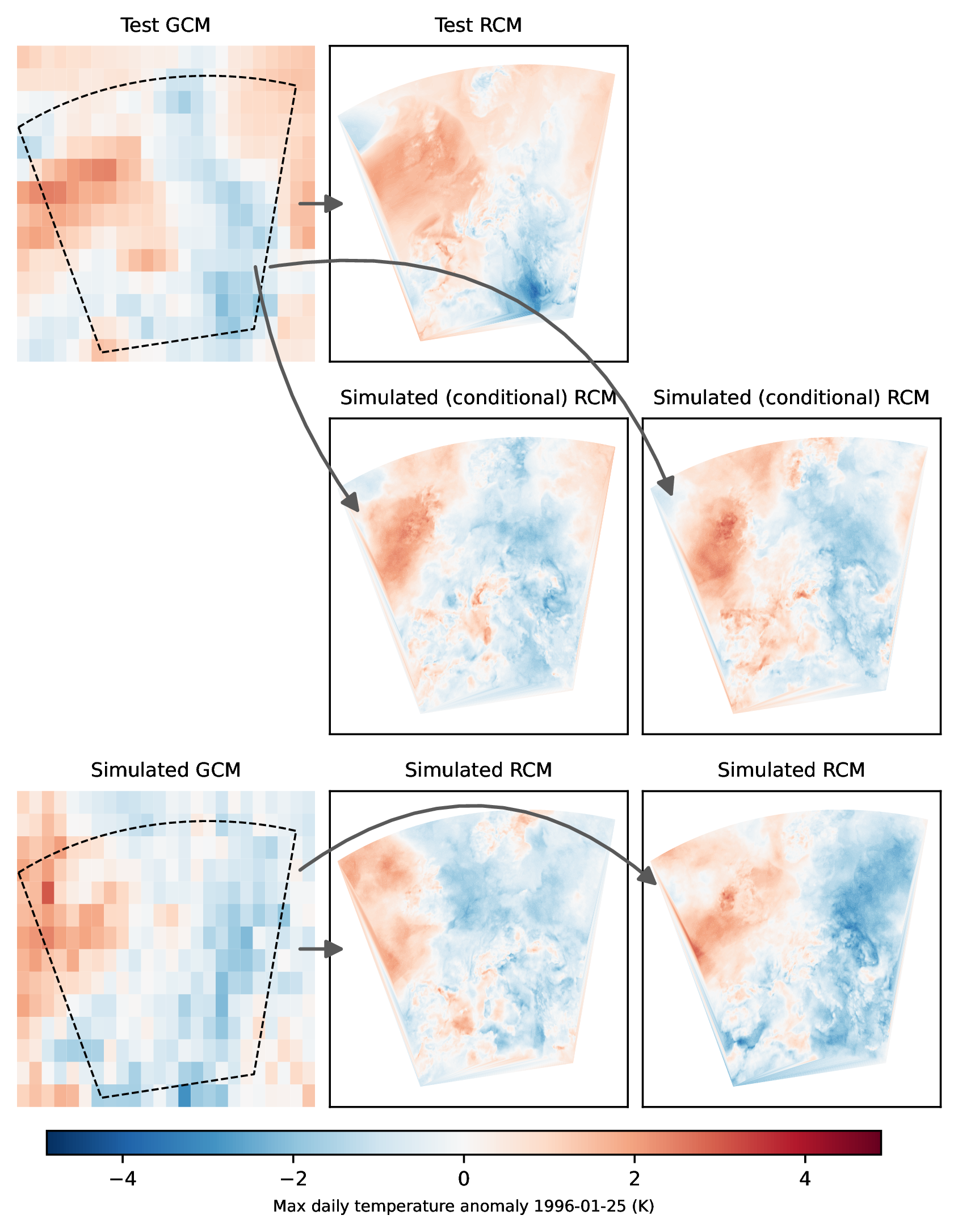}
    \caption{\textbf{Our MS BTM approach can capture the joint and conditional spatial structure encoded by GCM-RCM climate models (Section~\ref{sec:climate}).} First row: Ensemble member of the maximum temperature anomalies for the CanESM2 ($N_1 = 336$) and CRM5 ($N_2 = 78{,}400$) pairings over Europe. Our MS BTM infers the $N_1 + N_2$ dimensional distribution of coarse and fine scaled features, including the $N_2$-dimensional conditional (on $N_1$ values) distribution, from $n = 40$ training samples. The second and third row show samples from our model, conditioning on lower-resolution (GCM) fields, where the arrows represent the conditioning structure. We see that samples from our fitted model look qualitatively similar to the actual (held-out) observations.}
    \label{fig:anomalies_climate}
\end{figure}

We consider maximum daily temperatures ($K$) on a longitude-latitude regional grid (over Europe) of size $N_1 = 24 \times 14 = 336$ for the CanESM2 output, and a grid of size $N_2 = 280 \times 280 = 78{,}400$ for the CRCM5 output. We define each location $\bs_{r,i}$ as the center of the corresponding grid cell, so that $\ell_{r,i}$ is computed based on distances between these centers. We obtain temperature anomalies by standardizing the data at each grid location to zero mean and unit variance. Examples are shown in Figures~\ref{fig:anomalies_samples} and the first row of Figure \ref{fig:anomalies_climate}. We limit our analysis to a single day, January 25th 1996, but extending the model to include temporal structure is straightforward (see brief discussion in Section \ref{sec:conclusions}). 

\begin{figure}[htpb]
    \centering
    \includegraphics[width=0.5\linewidth]{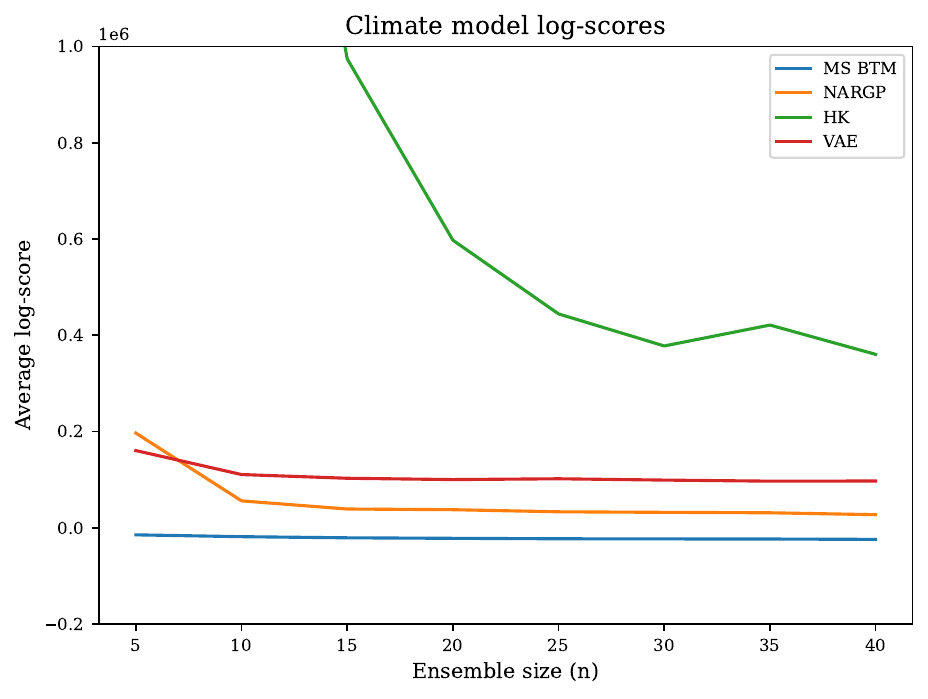}
    \caption{\textbf{The multi-scale BTM outperformed all other methods for all ensemble sizes on our GCM-RCM setup}. NARGP and VAE had very similar performance, and HK struggled with low ensemble size.}
    \label{fig:ls_climate}
\end{figure}

To evaluate the models, we compare the log-scores of the predictive distributions on $10$ held-out test samples, while varying the number $n$ of training samples; the results are shown in Figure \ref{fig:ls_climate}. We also plot  (conditional and unconditional) GCM-RCM sample pairs in Figure \ref{fig:anomalies_climate}. Again, the MS BTM is the best model in terms of average log-score for all ensemble sizes. The VAE is the second best model, but it appears that generating informative encodings would require a huge number of training samples, which is prohibitive in this scenario given the costly nature of RCM simulations. 

\section{Conclusions \label{sec:conclusions}}

We have proposed a Bayesian approach for learning non-Gaussian, high-dimensional distributions of multi-scale spatial fields based on a relatively small number of training samples. We do so via transport maps, whose components are modeled using autoregressive Gaussian processes, which have a spatially informed parametrization that only requires a small number of parameters to be fitted. The autoregressive nature of our map allows us to naturally learn the conditional distributions of higher resolution given lower resolution, which, for example, enables cheap sampling of (otherwise costly) high-resolution fields given only (relatively cheap) low-resolution observations. Our Bayesian approach probabilistically regularizes the model, resulting in sparse maps that are scalable in high dimensions. The method can be fitted quickly and was highly accurate in our numerical comparisons, outperforming all considered competing methods. It is therefore a promising approach for many spatial multi-scale surrogate tasks, especially when low-scale models are much cheaper to run than their high-scale counterparts. Potentially, deep models such as the VAE could outperform the MS BTM method in data-rich regimes, where much more training samples are available than in the settings considered here.

For future work, it is straightforward to combine the proposed multi-scale approach with existing spatio-temporal %
\citep{lei2026scalable} and multiple-process \citep{Wiemann2023BayesianFields} extensions of the single-scale BTM method, which rely on dimension-expanded input spaces. We can also model the marginal distributions more flexibly \citep{Brachem2026}, consider parametric shrinkage for scenarios where we have only one or very few training samples \citep{chakraborty2024}, encode explicit physical constraints between resolutions (such as conservation of mass or energy), or extend the MS-BTM from a Markovian chain of dependence between resolutions to a sparse directed acyclic graph over diverse sources of information. Possible applications include multi-scale climate model emulation (especially as surrogates for the GCM-RCM paradigm presented in Section \ref{sec:climate}), RCM parameter calibration given GCM outputs, or attribution of regional climate-change-related extremes.

\footnotesize
\appendix

\section*{Acknowledgments}

The authors were partially supported by NASA's Advanced Information Systems Technology Program (AIST-21). ASC's and MK's research was also partially supported by National Science Foundation (NSF) Grant DMS--1953005/2433548, by industry members of the Center for Interdisciplinary Research on Convective Storms (CIRCS) and National Science Foundation award numbers 2517152 and 2517615, and by the Office of the Vice Chancellor for Research and Graduate Education at the University of Wisconsin--Madison with funding from the Wisconsin Alumni Research Foundation. We would like to thank Daniel Wright for helpful comments and discussions. Part of this research was performed while ASC was visiting the Institute for Mathematical and Statistical Innovation (IMSI), which is supported by the National Science Foundation (Grant DMS-2425650).

\section*{Conflict of Interest Statement}

The authors report that there are no competing interests to declare.

\section*{Data Availability Statement}

The data that support the findings of this study are openly available in \url{https://crd-data-donnees-rdc.ec.gc.ca/CCCMA/products/CanSISE/output/CCCma/CanESM2} (GCM) and \url{https://climex-data.srv.lrz.de/Public/CanESM2_driven_50_members/tasmax} (RCM). Scripts to download the relevant data from these repositories are available in the repository\url{https://github.com/katzfuss-group/batram/tree/mf/tests/data}

\section{Competitors: Implementation details}\label{sec:app:competitors}

\paragraph{Matérn}

If we assume that the averages of higher scales correspond to observations in lower scales, as in \citet{ma2019b}, we can write
\[
\by_{r-1} = \bA \by_{r} + \by_{r}^{h},
\]
where $\bA$ is an averaging matrix, and $\by_{r}^{h}$ is a vector containing one hold-out fine-scale value from each coarse grid cell (and so $\by_{r}$ only contains the fine-scale values that are not held-out). Note that, in practice, we do not have access to $\bA$, nor do the relationships between the scales ought to be linear. We can model all resolutions jointly by assuming a GP prior with Matérn covariance on $\by_R$ (with left out points at all scales except the lowest, to obtain non-singular joint covariances for all scales), and finding parameters via second-order optimization. The conditional distributions in this scenario will be given by the usual conditional multivariate Gaussian formulas, and the left-out observations can be filled in deterministically. 

\paragraph{Hierarchical kriging (HK)}

For $R$ scales, \cite{han2012} propose setting the prior on the first scale as
\begin{equation*}
    y_{1}(\bs_1) \sim \GP(0, K_1(\bs_1, \bs_1^\prime)),
\end{equation*}
and, for the subsequent scales, they recursively set
\begin{equation*}
    y_{r}(\bs_r) \sim \GP(\beta_{0,r} \hat{y}_{r-1}(\bs_{r}), K_r(\bs_r, \bs_r^\prime)),
\end{equation*}
where $\beta_{0,r}$ is a scaling factor which measures correlation between the $r$-th and $r-1$-th resolutions, and $\hat{y}_{r-1}(\bs_{r})$ is the posterior mean of the GP at the $r-1$-th scale evaluated at the locations of the $r$-th scale. They use a spline correlation function \citep{lophaven2002dace} for their kernels $K_1, \dots, K_R$, but in our implementation we choose a more common Matérn kernel, due to the spatial nature of our data. We chose a smoothness of $\nu = 0.5$, which we found resulted in better log-scores. Also, they find their parameter estimates using genetic algorithms, but we fit them via SGD with auto-differentiation. 

One key assumption of the paper is that they only observe one spatial field per resolution. In our settings, we observe $n$ spatial fields per resolution, so we get $\beta_{0,r,1}, \dots, \beta_{0,r,N_r}$ estimates of the correlations between spatial fields at different scales for each resolution. In order to accommodate for these differing prior means, we normalize each field at resolution $r$, $\by_{r}^{(i)}$, using the corresponding $\beta_{0,r,i} \hat{y}_{r-1}(\bs_r)$ as the average of the field, and we put a prior mean at zero at each scale. 

For the climate-model application in Section~\ref{sec:climate}, where the number of locations in higher resolution is large, we fit the lower-resolution (GCM data) using an exact Matérn kernel, but for the RCM outputs we use a variational strategy to approximate the posterior with interpolation kernels \citep{wilson2015kernel} and inducing points \citep{hensman2013gaussian}.

\paragraph{Non-linear auto-regressive GP (NARGP)}
\cite{perdikaris2017} retain the standard GP prior on the
coarsest level,
\begin{equation*}
    y_{1}(\bs_1) \sim \GP\!\bigl(0,\; K_{1}(\bs_1,\bs_1^\prime)\bigr),
\end{equation*}
but model every subsequent-scale by augmenting the spatial input with
the posterior mean prediction of the previous level.  Let
$\hat{y}_{r-1}(\bs_{r})$ denote the GP posterior mean at resolution
$r\!-\!1$ evaluated at the locations of scale $r$ and define the
augmented input
$\tilde{\bx}_{r} := \bigl[\hat{y}_{r-1}(\bs_{r})\;,\;\bs_{r}^{\!\top}\bigr]^{\!\top}\!\in\!\mathbb{R}^{d+1}$.
The $r$-th scale is then assigned the GP prior
\begin{equation*}
    \by_{r}(\bs_{r})
    \;\sim\;
    \GP\!\bigl(0,\; K_{r}(\tilde{\bx}_{r},\tilde{\bx}_{r}^{\prime})\bigr),
    \qquad\quad r = 2,\dots,R,
\end{equation*}
with a kernel that factorizes into a product-sum form
\begin{equation*}
    K_{r}(\tilde{\bx},\tilde{\bx}^{\prime})
    \;=\;
    \,k_a\!\bigl(\hat{\by}_{r-1}(\bs_{r}),\hat{\by}_{r-1}(\bs_{r}^\prime)\bigr)\,
                  k_x\!\bigl(\bs_r,\bs_r^{\prime}\bigr)
    \;+\;
    k_x\!\bigl(\bs_r,\bs_r^{\prime}\bigr),
\end{equation*}
where $k_a$ acts on the low-scale prediction (capturing possibly
non-linear relations across scales) and $k_x$ is a spatial kernel.
We again use an exponential Matérn kernel ($\nu\!=\!0.5$) for
$k_x$ and $k_a$, as it yielded lower log-scores on our climate data than smoother choices. Each kernel has its own ARD weights for each dimension.

For $R > 2$, since the posterior distribution at all resolutions except $r = 1$ is no longer Gaussian, we propagate uncertainty along each recursive step, and approximate the posterior likelihood using Monte Carlo integration, as \cite{perdikaris2017}.

For the climate example, the posterior for the higher-scale GP is too costly to learn, so we use the same variational strategies as used in HK. 

\paragraph{$R$-scale variational autoencoder (VAE)}

Inspired by \citet{cheng2024bi}, for a multi-scale variational autoencoder we treat the collection of spatial fields observed at increasing resolutions as a single hierarchy generated from shared latent structure. We introduce an independent latent vector $\bz_r$ for each resolution, which is drawn from a standard-normal prior, and all of them are concatenated into a global latent vector $\bz = (\bz_1^\top, \dots, \bz_R^\top)^\top$. 

This latent vector is passed through convolutional layers to learn a common feature map, and then we have $R$ decoder heads that up-samples the shared feature map so that it matches the target grid size at the $r$-th resolution, which outputs the mean of a diagonal Gaussian likelihood for $\by_r$. A learnable variance $\sigma^2_r$ completes the generative model
\begin{equation*}
    \by_r|\bz \sim \normal(f_\theta^{(r)}(\bz), \sigma_r^2 \bI_{N_r}).
\end{equation*}

Inference is amortized with $R$ separate convolutional encoders, which maps its input field to the mean and log-variance of a diagonal Gaussian posterior $q_\phi(\bz_r|\by_r)$. Under a mean-field assumption \citep{Blei2017}, the global variational distribution factorizes as the product of these $R$ Gaussians. This leads to a evidence lower bound (ELBO) that is a sum over scales of reconstruction terms (one log-likelihood per scale) minus a sum of KL divergences that regularize every scale, with some $\beta_r \in [0,1]$ terms that allows us to tune how hard we push towards the prior \citep{higgins2017beta}. We optimize the resulting objective with the usual reparameterization trick \citep{Kingma2014,rezende2015} and ADAM, using mini-batches that carry observations from all resolutions. 

To estimate predictive performance comparable with our methods here, we approximate the marginal log-likelihood of a multi-resolution observation via importance-weighted sampling \citep{burda2015importance}. For every held-out test set member, we draw a few thousand latent samples from the variational posterior and evaluate their joint likelihood across all scales, and average the resulting importance weights. 

In practice, we use smaller latent dimensions at lower-resolutions fields and larger ones at higher resolutions, together with ELU activations and batch normalization. Thus, the model learns a single coherent latent representation that explains all scales simultaneously, but allows each resolution to express its own detail through their dedicated decoder.

\bibliographystyle{apalike}
\bibliography{mendeley,additionalrefs}

@article{katzfuss2023,
author = {Matthias Katzfuss and Florian Sch{{\"a}}fer},
title = {Scalable Bayesian Transport Maps for High-Dimensional Non-Gaussian Spatial Fields},
journal = {Journal of the American Statistical Association},
volume = {0},
number = {0},
pages = {1-15},
year = {2023},
publisher = {Taylor & Francis},
doi = {10.1080/01621459.2023.2197158},


URL = { 
    
        https://doi.org/10.1080/01621459.2023.2197158
    
    

},
eprint = { 
    
        https://doi.org/10.1080/01621459.2023.2197158
    
    

}

}

@article{goyal2017accurate,
  title={Accurate, large minibatch sgd: Training imagenet in 1 hour},
  author={Goyal, Priya and Doll{\'a}r, Piotr and Girshick, Ross and Noordhuis, Pieter and Wesolowski, Lukasz and Kyrola, Aapo and Tulloch, Andrew and Jia, Yangqing and He, Kaiming},
  journal={arXiv preprint arXiv:1706.02677},
  year={2017}
}

@article{giorgi1989,
  title={The climatological skill of a regional model over complex terrain},
  author={Giorgi, Filippo and Bates, Gary T},
  journal={Monthly Weather Review},
  volume={117},
  number={11},
  pages={2325--2347},
  year={1989}
}

@article{coppola2021,
  title={Non-Hydrostatic RegCM4 (RegCM4-NH): model description and case studies over multiple domains},
  author={Coppola, Erika and Stocchi, Paolo and Pichelli, Emanuela and Torres Alavez, Jose Abraham and Glazer, Russell and Giuliani, Graziano and Di Sante, Fabio and Nogherotto, Rita and Giorgi, Filippo},
  journal={Geoscientific Model Development Discussions},
  volume={2021},
  pages={1--25},
  year={2021},
  publisher={G{\"o}ttingen, Germany}
}

@article{giorgi2023,
  title={The Fifth Generation Regional Climate Modeling System, RegCM5: Description and Illustrative Examples at Parameterized Convection and Convection-Permitting Resolutions},
  author={Giorgi, Filippo and Coppola, Erika and Giuliani, Graziano and Ciarlo, James M and Pichelli, Emanuela and Nogherotto, Rita and Raffaele, Francesca and Malguzzi, Piero and Davolio, Silvio and Stocchi, Paolo and others},
  journal={Journal of Geophysical Research: Atmospheres},
  volume={128},
  number={6},
  pages={e2022JD038199},
  year={2023},
  publisher={Wiley Online Library}
}

@article{ma2022,
  title={Multifidelity computer model emulation with high-dimensional output: An application to storm surge},
  author={Ma, Pulong and Karagiannis, Georgios and Konomi, Bledar A and Asher, Taylor G and Toro, Gabriel R and Cox, Andrew T},
  journal={Journal of the Royal Statistical Society Series C: Applied Statistics},
  volume={71},
  number={4},
  pages={861--883},
  year={2022},
  publisher={Oxford University Press}
}

@misc{mearns2017,
  title={The NA-CORDEX dataset, version 1.0. NCAR climate data gateway, boulder CO},
  author={Mearns, L and McGinnis, S and Korytina, D and Arritt, R and Biner, S and Bukovsky, M and Chang, HI and Christensen, O and Herzmann, D and Jiao, Yanjun and others},
  year={2017}
}

@article{han2012,
  title={Hierarchical kriging model for variable-fidelity surrogate modeling},
  author={Han, Zhong-Hua and G{\"o}rtz, Stefan},
  journal={AIAA journal},
  volume={50},
  number={9},
  pages={1885--1896},
  year={2012}
}

@article{kennedy2000,
  title={Predicting the output from a complex computer code when fast approximations are available},
  author={Kennedy, Marc C and O'Hagan, Anthony},
  journal={Biometrika},
  volume={87},
  number={1},
  pages={1--13},
  year={2000},
  publisher={Oxford University Press}
}

@article{le2014,
  title={Recursive co-kriging model for design of computer experiments with multiple levels of fidelity},
  author={Le Gratiet, Loic and Garnier, Josselin},
  journal={International Journal for Uncertainty Quantification},
  volume={4},
  number={5},
  year={2014},
  publisher={Begel House Inc.}
}

@article{cheng2023,
  title={Recursive nearest neighbor co-kriging models for big multi-fidelity spatial data sets},
  author={Cheng, Si and Konomi, Bledar A and Karagiannis, Georgios and Kang, Emily L},
  journal={Environmetrics},
  pages={e2844},
  year={2023},
  publisher={Wiley Online Library}
}

@InProceedings{niu2024,
  title = 	 {Multi-Fidelity Residual Neural Processes for Scalable Surrogate Modeling},
  author =       {Niu, Ruijia and Wu, Dongxia and Kim, Kai and Ma, Yian and Watson-Parris, Duncan and Yu, Rose},
  booktitle = 	 {Proceedings of the 41st International Conference on Machine Learning},
  pages = 	 {38381--38394},
  year = 	 {2024},
  editor = 	 {Salakhutdinov, Ruslan and Kolter, Zico and Heller, Katherine and Weller, Adrian and Oliver, Nuria and Scarlett, Jonathan and Berkenkamp, Felix},
  volume = 	 {235},
  series = 	 {Proceedings of Machine Learning Research},
  month = 	 {21--27 Jul},
  publisher =    {PMLR},
  url = 	 {https://proceedings.mlr.press/v235/niu24d.html}
}

@article{buster2024,
  title={High-resolution meteorology with climate change impacts from global climate model data using generative machine learning},
  author={Buster, Grant and Benton, Brandon N and Glaws, Andrew and King, Ryan N},
  journal={Nature Energy},
  pages={1--13},
  year={2024},
  publisher={Nature Publishing Group UK London}
}

@article{schafer2021,
author = {Sch\"{a}fer, Florian and Sullivan, T. J. and Owhadi, Houman},
title = {Compression, Inversion, and Approximate PCA of Dense Kernel Matrices at Near-Linear Computational Complexity},
journal = {Multiscale Modeling \& Simulation},
volume = {19},
number = {2},
pages = {688-730},
year = {2021},
doi = {10.1137/19M129526X},

URL = { 
    
        https://doi.org/10.1137/19M129526X
    
    

},
eprint = { 
    
        https://doi.org/10.1137/19M129526X
    
    

}
}

@article{yuUpperTailPrecipitation2020,
author = {Yu, Guo and Wright, Daniel B. and Li, Zhe},
title = {The Upper Tail of Precipitation in Convection-Permitting Regional Climate Models and Their Utility in Nonstationary Rainfall and Flood Frequency Analysis},
journal = {Earth's Future},
volume = {8},
number = {10},
pages = {e2020EF001613},
doi = {https://doi.org/10.1029/2020EF001613},
url = {https://agupubs.onlinelibrary.wiley.com/doi/abs/10.1029/2020EF001613},
eprint = {https://agupubs.onlinelibrary.wiley.com/doi/pdf/10.1029/2020EF001613},
note = {e2020EF001613 2020EF001613},
year = {2020}
}

@article{mardani2024residual,
  author  = {Mardani, Morteza and Brenowitz, Noah and Cohen, Yair and Pathak, Jaideep and Chen, Chieh-Yu and Liu, Cheng-Chin and Vahdat, Arash and Nabian, Mohammad Amin and Ge, Tao and Subramaniam, Akshay and Kashinath, Karthik and Kautz, Jan and Pritchard, Mike},
  title   = {Residual corrective diffusion modeling for km-scale atmospheric downscaling},
  journal = {Communications Earth \& Environment},
  year    = {2025},
  volume  = {6},
  number  = {1},
  pages   = {124},
  doi     = {10.1038/s43247-025-02042-5},
  url     = {https://doi.org/10.1038/s43247-025-02042-5}
}

@article{perdikaris2017,
  title={Nonlinear information fusion algorithms for data-efficient multi-fidelity modelling},
  author={Perdikaris, Paris and Raissi, Maziar and Damianou, Andreas and Lawrence, Neil D and Karniadakis, George Em},
  journal={Proceedings of the Royal Society A: Mathematical, Physical and Engineering Sciences},
  volume={473},
  number={2198},
  pages={20160751},
  year={2017},
  publisher={The Royal Society Publishing}
}

@inproceedings{rezende2015,
  title={Variational inference with normalizing flows},
  author={Rezende, Danilo and Mohamed, Shakir},
  booktitle={International conference on machine learning},
  pages={1530--1538},
  year={2015},
  organization={PMLR}
}

@inproceedings{wu2022multi,
  title={Multi-fidelity hierarchical neural processes},
  author={Wu, Dongxia and Chinazzi, Matteo and Vespignani, Alessandro and Ma, Yi-An and Yu, Rose},
  booktitle={Proceedings of the 28th ACM SIGKDD Conference on Knowledge Discovery and Data Mining},
  pages={2029--2038},
  year={2022}
}

@article{chakraborty2024,
  title={Learning non-Gaussian spatial distributions via Bayesian transport maps with parametric shrinkage},
  author={Chakraborty, Anirban and Katzfuss, Matthias},
  journal={Journal of Agricultural, Biological and Environmental Statistics},
  pages={1--19},
  year={2025},
  publisher={Springer}
}

@techreport{lophaven2002dace,
title = "DACE - A Matlab Kriging Toolbox, Version 2.0",
author = "Lophaven, {S{\o}ren Nymand} and Nielsen, {Hans Bruun} and Jacob S{\o}ndergaard",
year = "2002",
language = "English",
institution = "Technical University of Denmark"
}

@article{ma2019b,
author = {Pulong Ma and Emily L. Kang and Amy J. Braverman and Hai M. Nguyen},
title = {Spatial Statistical Downscaling for Constructing High-Resolution Nature Runs in Global Observing System Simulation Experiments},
journal = {Technometrics},
volume = {61},
number = {3},
pages = {322-340},
year = {2019},
publisher = {Taylor & Francis},
doi = {10.1080/00401706.2018.1524791}
}

@inproceedings{chattopadhyay2023long,
  title={Long-term instability of deep learning-based digital twins of the climate system: Cause and solution},
  author={Chattopadhyay, Ashesh and Hassanzadeh, Pedram},
  booktitle={APS March Meeting Abstracts},
  volume={2023},
  pages={W53--004},
  year={2023}
}

@article{baptista2024representation,
  title={On the representation and learning of monotone triangular transport maps},
  author={Baptista, Ricardo and Marzouk, Youssef and Zahm, Olivier},
  journal={Foundations of Computational Mathematics},
  volume={24},
  number={6},
  pages={2063--2108},
  year={2024},
  publisher={Springer}
}

@inproceedings{irons2022triangular,
  title={Triangular flows for generative modeling: Statistical consistency, smoothness classes, and fast rates},
  author={Irons, Nicholas J and Scetbon, Meyer and Pal, Soumik and Harchaoui, Zaid},
  booktitle={International Conference on Artificial Intelligence and Statistics},
  pages={10161--10195},
  year={2022},
  organization={PMLR}
}

@article{wang2022minimax,
  title={On minimax density estimation via measure transport},
  author={Wang, Sven and Marzouk, Youssef},
  journal={arXiv preprint arXiv:2207.10231},
  year={2022}
}

@article{chen2025sparse,
  title={Sparse Cholesky factorization for solving nonlinear PDEs via Gaussian processes},
  author={Chen, Yifan and Owhadi, Houman and Sch{\"a}fer, Florian},
  journal={Mathematics of Computation},
  volume={94},
  number={353},
  pages={1235--1280},
  year={2025}
}

@article{ekanayaka2025multivariate,
  title={A multivariate spatial statistical model for statistical downscaling of sea surface temperature in the Great Barrier Reef region},
  author={Ekanayaka, Ayesha and Kang, Emily L and Braverman, Amy and Kalmus, Peter},
  journal={Journal of the Royal Statistical Society Series C: Applied Statistics},
  pages={qlaf019},
  year={2025},
  publisher={Oxford University Press UK}
}

@article{cheng2024bi,
  title={Bi-fidelity variational auto-encoder for uncertainty quantification},
  author={Cheng, Nuojin and Malik, Osman Asif and De, Subhayan and Becker, Stephen and Doostan, Alireza},
  journal={Computer Methods in Applied Mechanics and Engineering},
  volume={421},
  pages={116793},
  year={2024},
  publisher={Elsevier}
}

@article{baptista2020conditional,
  title={Conditional Sampling with Monotone GANs: From Generative Models to Likelihood-Free Inference},
  author={Ricardo Baptista and Bamdad Hosseini and Nikola B. Kovachki and Youssef M. Marzouk},
  journal={SIAM/ASA J. Uncertain. Quantification},
  year={2020},
  volume={12},
  pages={868-900},
  url={https://api.semanticscholar.org/CorpusID:259088624}
}

@book{adams2003sobolev,
  title={Sobolev spaces},
  author={Adams, Robert A and Fournier, John JF},
  volume={140},
  year={2003},
  publisher={Elsevier}
}

@article{chen2024precision,
author = {Chen, Jiaheng and Sanz-Alonso, Daniel},
title = {Precision and Cholesky Factor Estimation for Gaussian Processes},
journal = {SIAM/ASA Journal on Uncertainty Quantification},
volume = {13},
number = {3},
pages = {1085-1115},
year = {2025},
doi = {10.1137/24M1717282},

URL = { 
    
        https://doi.org/10.1137/24M1717282
    
    

},
eprint = { 
    
        https://doi.org/10.1137/24M1717282
    
    

}
}

@inproceedings{wilson2015kernel,
  title={Kernel interpolation for scalable structured Gaussian processes (KISS-GP)},
  author={Wilson, Andrew and Nickisch, Hannes},
  booktitle={International conference on machine learning},
  pages={1775--1784},
  year={2015},
  organization={PMLR}
}

@inproceedings{hensman2013gaussian,
author = {Hensman, James and Fusi, Nicol\`{o} and Lawrence, Neil D.},
title = {Gaussian processes for Big data},
year = {2013},
publisher = {AUAI Press},
address = {Arlington, Virginia, USA},
booktitle = {Proceedings of the Twenty-Ninth Conference on Uncertainty in Artificial Intelligence},
pages = {282–290},
numpages = {9},
location = {Bellevue, WA},
series = {UAI'13}
}

@inproceedings{higgins2017beta,
  title={beta-vae: Learning basic visual concepts with a constrained variational framework},
  author={Higgins, Irina and Matthey, Loic and Pal, Arka and Burgess, Christopher and Glorot, Xavier and Botvinick, Matthew and Mohamed, Shakir and Lerchner, Alexander},
  booktitle={International conference on learning representations},
  year={2017}
}

@article{burda2015importance,
  title={Importance weighted autoencoders},
  author={Burda, Yuri and Grosse, Roger and Salakhutdinov, Ruslan},
  journal={arXiv preprint arXiv:1509.00519},
  year={2015}
}

@article{kay2015community,
  title={The Community Earth System Model (CESM) large ensemble project: A community resource for studying climate change in the presence of internal climate variability},
  author={Kay, Jennifer E and Deser, Clara and Phillips, A and Mai, A and Hannay, Cecile and Strand, Gary and Arblaster, Julie Michelle and Bates, SC and Danabasoglu, Gokhan and Edwards, James and others},
  journal={Bulletin of the American Meteorological Society},
  volume={96},
  number={8},
  pages={1333--1349},
  year={2015}
}

@article{xu2019dynamical,
  title={Dynamical downscaling of regional climate: A review of methods and limitations},
  author={Xu, Zhongfeng and Han, Ying and Yang, Zongliang},
  journal={Science China Earth Sciences},
  volume={62},
  number={2},
  pages={365--375},
  year={2019},
  publisher={Springer}
}

@article{schneider2024opinion,
  title={Opinion: Optimizing climate models with process knowledge, resolution, and artificial intelligence},
  author={Schneider, Tapio and Leung, L Ruby and Wills, Robert CJ},
  journal={Atmospheric Chemistry and Physics},
  volume={24},
  number={12},
  pages={7041--7062},
  year={2024},
  publisher={Copernicus Publications G{\"o}ttingen, Germany}
}

@article{kushner2018canadian,
  title={Canadian snow and sea ice: assessment of snow, sea ice, and related climate processes in Canada's Earth system model and climate-prediction system},
  author={Kushner, Paul J and Mudryk, Lawrence R and Merryfield, William and Ambadan, Jaison T and Berg, Aaron and Bichet, Ad{\'e}line and Brown, Ross and Derksen, Chris and D{\'e}ry, Stephen J and Dirkson, Arlan and others},
  journal={The Cryosphere},
  volume={12},
  number={4},
  pages={1137--1156},
  year={2018},
  publisher={Copernicus GmbH}
}

@article{kirchmeier2017attribution,
  title={Attribution of extreme events in Arctic sea ice extent},
  author={Kirchmeier-Young, Megan C and Zwiers, Francis W and Gillett, Nathan P},
  journal={Journal of Climate},
  volume={30},
  number={2},
  pages={553--571},
  year={2017}
}

@article{leduc2019climex,
  title={The ClimEx project: A 50-member ensemble of climate change projections at 12-km resolution over Europe and northeastern North America with the Canadian Regional Climate Model (CRCM5)},
  author={Leduc, Martin and Mailhot, Alain and Frigon, Anne and Martel, Jean-Luc and Ludwig, Ralf and Brietzke, Gilbert B and Gigu{\`e}re, Michel and Brissette, Fran{\c{c}}ois and Turcotte, Richard and Braun, Marco and others},
  journal={Journal of Applied Meteorology and Climatology},
  volume={58},
  number={4},
  pages={663--693},
  year={2019}
}

@article{chylek2011observed,
  title={Observed and model simulated 20th century Arctic temperature variability: Canadian earth system model CanESM2},
  author={Chylek, P and Li, J and Dubey, MK and Wang, M and Lesins, GJAC},
  journal={Atmospheric Chemistry and Physics Discussions},
  volume={11},
  number={8},
  pages={22893--22907},
  year={2011},
  publisher={Copernicus GmbH}
}

@article{martynov2013reanalysis,
  title={Reanalysis-driven climate simulation over CORDEX North America domain using the Canadian Regional Climate Model, version 5: model performance evaluation},
  author={Martynov, Andrey and Laprise, Ren{\'e} and Sushama, Laxmi and Winger, Katja and {\v{S}}eparovi{\'c}, L and Dugas, B},
  journal={Climate dynamics},
  volume={41},
  number={11},
  pages={2973--3005},
  year={2013},
  publisher={Springer}
}

@article{chou1994multiscale,
  title={Multiscale recursive estimation, data fusion, and regularization},
  author={Chou, Kenneth C and Willsky, Alan S and Benveniste, Albert},
  journal={IEEE transactions on Automatic Control},
  volume={39},
  number={3},
  pages={464--478},
  year={1994},
  publisher={IEEE}
}

@article{ramasamy2013fusion,
    author = {Ramasamy, Suresh K. and Raja, Jayaraman and Boudreau, Brian D.},
    title = {Data Fusion Strategy for Multiscale Surface Measurements},
    journal = {Journal of Micro and Nano-Manufacturing},
    volume = {1},
    number = {1},
    pages = {011004},
    year = {2013},
    month = {03},
    issn = {2166-0468},
    doi = {10.1115/1.4023755},
    url = {https://doi.org/10.1115/1.4023755},
    eprint = {https://asmedigitalcollection.asme.org/micronanomanufacturing/article-pdf/1/1/011004/6062278/jmnm_1_1_011004.pdf},
}

@inproceedings{gonzalez2023multi,
  title={Multi-variable hard physical constraints for climate model downscaling},
  author={Gonz{\'a}lez-Abad, Jose and Hern{\'a}ndez-Garc{\'\i}a, {\'A}lex and Harder, Paula and Rolnick, David and Guti{\'e}rrez, Jos{\'e} Manuel},
  booktitle={Proceedings of the aaai symposium series},
  volume={2},
  pages={62--67},
  year={2023}
}

@Article{aich2024conditional,
AUTHOR = {Aich, M. and Hess, P. and Pan, B. and Bathiany, S. and Huang, Y. and Boers, N.},
TITLE = {Conditional diffusion models for downscaling \& bias correction of Earth system model precipitation},
JOURNAL = {EGUsphere},
VOLUME = {2025},
YEAR = {2025},
PAGES = {1--21},
URL = {https://egusphere.copernicus.org/preprints/2025/egusphere-2025-2646/},
DOI = {10.5194/egusphere-2025-2646}
}

@article{singh2023deep,
  title={Deep learning and data fusion to estimate surface soil moisture from multi-sensor satellite images},
  author={Singh, Abhilash and Gaurav, Kumar},
  journal={Scientific Reports},
  volume={13},
  number={1},
  pages={2251},
  year={2023},
  publisher={Nature Publishing Group UK London}
}

@article{zhang2012urban,
author = {Zhang, Yang and Karamchandani, Prakash and Glotfelty, Tim and Streets, David G. and Grell, Georg and Nenes, Athanasios and Yu, Fangqun and Bennartz, Ralf},
title = {Development and initial application of the global-through-urban weather research and forecasting model with chemistry (GU-WRF/Chem)},
journal = {Journal of Geophysical Research: Atmospheres},
volume = {117},
number = {D20},
pages = {},
doi = {https://doi.org/10.1029/2012JD017966},
url = {https://agupubs.onlinelibrary.wiley.com/doi/abs/10.1029/2012JD017966},
eprint = {https://agupubs.onlinelibrary.wiley.com/doi/pdf/10.1029/2012JD017966},
year = {2012}
}

@article{chen2009regional,
  title={Regional CO pollution and export in China simulated by the high-resolution nested-grid GEOS-Chem model},
  author={Chen, Dan and Wang, Yuxuan and McElroy, Michael Brendon and He, Kebin and Yantosca, Robert M and Le Sager, Philippe},
  journal={Atmospheric Chemistry and Physics},
  volume={9},
  number={11},
  pages={3825--3839},
  year={2009},
  publisher={Copernicus GmbH}
}

@book{maraun2018statistical,
  title={Statistical downscaling and bias correction for climate research},
  author={Maraun, Douglas and Widmann, Martin},
  year={2018},
  publisher={Cambridge University Press}
}

@article{maraun2010downscaling,
author = {Maraun, D. and Wetterhall, F. and Ireson, A. M. and Chandler, R. E. and Kendon, E. J. and Widmann, M. and Brienen, S. and Rust, H. W. and Sauter, T. and Themeßl, M. and Venema, V. K. C. and Chun, K. P. and Goodess, C. M. and Jones, R. G. and Onof, C. and Vrac, M. and Thiele-Eich, I.},
title = {Precipitation downscaling under climate change: Recent developments to bridge the gap between dynamical models and the end user},
journal = {Reviews of Geophysics},
volume = {48},
number = {3},
pages = {},
doi = {https://doi.org/10.1029/2009RG000314},
url = {https://agupubs.onlinelibrary.wiley.com/doi/abs/10.1029/2009RG000314},
eprint = {https://agupubs.onlinelibrary.wiley.com/doi/pdf/10.1029/2009RG000314},
year = {2010}
}

@article{pan2025mesmer,
  title={MESMER-RCM: A Probabilistic Climate Emulator for Regional Warming Projections},
  author={Pan, Hao and Gudmundsson, Lukas and Hauser, Mathias and Schwaab, Jonas and Quilcaille, Yann and Seneviratne, Sonia I},
  journal={EGUsphere},
  volume={2025},
  pages={1--13},
  year={2025},
  publisher={Copernicus Publications G{\"o}ttingen, Germany}
}

@article{Brachem2026,
    title = {{Data-efficient generative modeling of non-Gaussian global climate fields via scalable composite transformations}},
    year = {2026},
    journal = {arXiv:2602.23311},
    author = {Brachem, Johannes and Wiemann, Paul F. V. and Katzfuss, Matthias},
    month = {2},
    arxivId = {2602.23311}
}

@article{lei2026scalable,
  title={Scalable generative modeling of non-Gaussian spatio-temporal fields via autoregressive Gaussian processes},
  author={Lei-Cramer, Carrie J and Cao, Jian and Katzfuss, Matthias},
  journal={arXiv preprint arXiv:2605.03152},
  year={2026}
}

@article{Gotway2002,
author = {Carol A Gotway and Linda J Young},
title = {Combining Incompatible Spatial Data},
journal = {Journal of the American Statistical Association},
volume = {97},
number = {458},
pages = {632--648},
year = {2002},
publisher = {Taylor \& Francis},
doi = {10.1198/016214502760047140},


URL = { 
    
        https://doi.org/10.1198/016214502760047140
    
    

},
eprint = { 
    
        https://doi.org/10.1198/016214502760047140
    
    

}

}

@article{gelfand2001,
    author = {Gelfand, Alan E. and Zhu, Li and Carlin, Bradley P.},
    title = {On the change of support problem for spatio-temporal data },
    journal = {Biostatistics},
    volume = {2},
    number = {1},
    pages = {31-45},
    year = {2001},
    month = {03},
    issn = {1465-4644},
    doi = {10.1093/biostatistics/2.1.31},
    url = {https://doi.org/10.1093/biostatistics/2.1.31},
    eprint = {https://academic.oup.com/biostatistics/article-pdf/2/1/31/654877/020031.pdf},
}

@article{Mearns2009,
    title = {{A regional climate change assessment program for North America}},
    year = {2009},
    journal = {Eos},
    author = {Mearns, L. O. and Gutowski, W. and Jones, R. and Leung, R. and Mcginnis, S. and Nunes, A. and Qian, Y.},
    number = {36},
    pages = {311--312},
    volume = {90},
    doi = {10.1029/2009EO360002},
    issn = {00963941}
}

@article{Berrocal2010,
    title = {{A Spatio-Temporal Downscaler for Output From Numerical Models}},
    year = {2010},
    journal = {Journal of Agricultural, Biological, and Environmental Statistics},
    author = {Berrocal, Veronica J. and Gelfand, Alan E. and Holland, David M.},
    number = {2},
    month = {1},
    pages = {176--197},
    volume = {15},
    url = {http://www.springerlink.com/index/10.1007/s13253-009-0004-z},
    doi = {10.1007/s13253-009-0004-z},
    issn = {1085-7117}
}

@inproceedings{Kingma2014,
    title = {{Auto-encoding variational Bayes}},
    year = {2014},
    booktitle = {International Conference on Learning Representations},
    author = {Kingma, Diederik P. and Welling, Max},
    arxivId = {1312.6114}
}

@article{Vecchia1988,
    title = {{Estimation and model identification for continuous spatial processes}},
    year = {1988},
    journal = {Journal of the Royal Statistical Society, Series B},
    author = {Vecchia, AV},
    number = {2},
    pages = {297--312},
    volume = {50},
    url = {http://www.jstor.org/stable/10.2307/2345768}
}

@article{Wiemann2023BayesianFields,
    title = {{Bayesian nonparametric generative modeling of large multivariate non-Gaussian spatial fields}},
    year = {2023},
    journal = {Journal of Agricultural, Biological and Environmental Statistics},
    author = {Wiemann, Paul F. V. and Katzfuss, Matthias},
    number = {4},
    month = {12},
    pages = {597--617},
    volume = {28},
    doi = {10.1007/s13253-023-00580-z},
    issn = {1085-7117}
}

@article{Guinness2016a,
    title = {{Permutation and grouping methods for sharpening Gaussian process approximations}},
    year = {2018},
    journal = {Technometrics},
    author = {Guinness, Joseph},
    number = {4},
    month = {10},
    pages = {415--429},
    volume = {60},
    url = {http://arxiv.org/abs/1609.05372 https://www.tandfonline.com/doi/full/10.1080/00401706.2018.1437476},
    doi = {10.1080/00401706.2018.1437476},
    issn = {0040-1706},
    arxivId = {1609.05372}
}

@article{Gneiting2014,
    title = {{Probabilistic forecasting}},
    year = {2014},
    journal = {Annual Review of Statistics and Its Application},
    author = {Gneiting, Tilmann and Katzfuss, Matthias},
    number = {1},
    month = {1},
    pages = {125--151},
    volume = {1},
    url = {http://www.annualreviews.org/doi/abs/10.1146/annurev-statistics-062713-085831},
    doi = {10.1146/annurev-statistics-062713-085831},
    issn = {2326-8298}
}

@incollection{Marzouk2016,
    title = {{Sampling via measure transport: An introduction}},
    year = {2016},
    booktitle = {Handbook of Uncertainty Quantification},
    author = {Marzouk, Youssef M. and Moselhy, Tarek and Parno, Matthew and Spantini, Alessio},
    editor = {Ghanem, R and Higdon, Dave and Owhadi, Houman},
    publisher = {Springer},
    isbn = {9783319123851},
    doi = {10.1007/978-3-319-12385-1}
}

@article{Schafer2020,
    title = {{Sparse Cholesky factorization by Kullback-Leibler minimization}},
    year = {2021},
    journal = {SIAM Journal on Scientific Computing},
    author = {Sch{\"{a}}fer, Florian and Katzfuss, Matthias and Owhadi, Houman},
    number = {3},
    pages = {A2019-A2046},
    volume = {43},
    doi = {10.1137/20M1336254},
    arxivId = {2004.14455}
}

@article{Stein2002,
    title = {{The screening effect in kriging}},
    year = {2002},
    journal = {Annals of Statistics},
    author = {Stein, Michael L.},
    number = {1},
    pages = {298--323},
    volume = {30},
    doi = {10.1214/aos/1015362194},
    issn = {0090-5364}
}

@article{Blei2017,
    title = {{Variational Inference: A Review for Statisticians}},
    year = {2017},
    journal = {Journal of the American Statistical Association},
    author = {Blei, David M. and Kucukelbir, Alp and McAuliffe, Jon D.},
    number = {518},
    pages = {859--877},
    volume = {112},
    doi = {10.1080/01621459.2017.1285773},
    issn = {1537274X},
    arxivId = {1601.00670}
}

@article{Stein2011,
    title = {{When does the screening effect hold?}},
    year = {2011},
    journal = {Annals of Statistics},
    author = {Stein, Michael L.},
    number = {6},
    month = {12},
    pages = {2795--2819},
    volume = {39},
    doi = {10.1214/11-AOS909},
    issn = {0090-5364}
}

\end{document}